\documentclass[11pt]{article}
\usepackage[T1]{fontenc}
\usepackage{lmodern,microtype,amsmath,amssymb,graphicx,booktabs,tabularx}
\usepackage[letterpaper,margin=0.86in,top=0.82in,bottom=.86in]{geometry}
\usepackage[numbers,sort&compress]{natbib}
\usepackage[hidelinks]{hyperref}
\hypersetup{pdftitle={Forecasting hidden-wall deformation under material ambiguity in elastic tubes},pdfauthor={\'I\~nigo Cirauqui-Viloria}}
\usepackage{caption}
\newcommand{\fig}[1]{\includegraphics[width=.94\linewidth]{figures/#1.pdf}}

\title{\bfseries Forecasting hidden-wall deformation under material ambiguity in elastic tubes}

\author{\'I\~nigo Cirauqui-Viloria$^{a,*}$\\ \small $^a$Universidad de Zaragoza, Pedro Cerbuna 12, 50009 Zaragoza, Spain\\
\small $^*$Corresponding author: \href{mailto:icirauqui@gmail.com}{icirauqui@gmail.com}\\
\small\href{https://orcid.org/0000-0003-1773-1937}{ORCID 0000-0003-1773-1937}}
\date{}
\begin{document}
\maketitle
\begin{abstract}

Can motion on one side of an elastic tube predict how the opposite wall will move under a new load when several stiffness distributions explain the observations? In numerical experiments, we use displacements at nine visible markers under two known loads to predict the hidden-wall displacement under a third prescribed load. The simulations include homogeneous tubes and tubes with a locally softer or stiffer wall patch (an inclusion).

A paired finite-element test identified the homogeneous wall response as the main source of forecast error after a change in element order. On 24 independent simulated tubes, replacing the linear-reference estimator with a homogeneous quadratic finite-element model reduced the mean relative hidden-wall error from 56.61\% to 3.51\%. The full quadratic-reference estimator reduced the total displacement error to 2.84\%. For the displacement change caused by adding the inclusion to the same tube under the same load, relative error exceeded 100\%. Simultaneous prediction bands covered 22 of 24 tubes, with a mean width above the prespecified target. For repeated predictions, a compliance-based graph filter reused the homogeneous factorization and achieved nearly the same blind forecast accuracy as converged local reanalysis under matched conditions.

These results demonstrate that accurate hidden-wall forecasts are possible under material ambiguity in the tested setting. The paired tests and independent confirmation show how to diagnose errors in the mechanical reference and recover forecast accuracy by improving the model of the homogeneous tube.

\end{abstract}

\section{Introduction}

A camera sees only one side of a deforming tube. If we know the applied
loads and measure how the visible side moves, can we predict how the
opposite wall will move under a new load? The answer depends on what the
visible motion tells us about the hidden wall. Two stiffness distributions
may explain nearly the same visible displacements. If they also predict
similar motion on the hidden side, the forecast can be accurate even when
the stiffness is uncertain. If their hidden predictions differ, the
ambiguity matters directly. This distinction between identifying parameters
and predicting a quantity of interest is central to goal-oriented Bayesian
inference~\citep{spantini}. It also changes what we ask of an additional
measurement. Active palpation and sensor design seek informative
observations~\citep{palpation,dino}. Here a useful observation distinguishes
models that predict different hidden motion under the new load.

Forecast accuracy also depends on the mechanical model. Even for a
homogeneous tube, an inaccurate finite-element solution can shift the
predictions for all candidate materials. We therefore separate the
displacement of the homogeneous tube from the change caused by a local
region of softer or stiffer material, which we call an inclusion. In our
numerical experiments, we observe displacements at nine markers under two
known loads. We use these readings to weight candidate responses and
predict the three-component displacement of the opposite wall under a
different prescribed load. The hidden displacements and material labels
are reserved for evaluation.

Our main contribution is a numerical assessment that distinguishes the
effect of material ambiguity from error in the mechanical reference.
We construct pairs of stiffness distributions with nearly indistinguishable
visible motion to bound the error of a hidden-wall forecast. In a paired
test, we hold the estimator fixed and change the finite-element order used
to generate the observations and targets. This identifies the homogeneous
wall response as the main source of the resulting forecast discrepancy.
An independent set of simulated tubes shows that rebuilding this reference
with quadratic elements recovers most of the lost forecast accuracy.
The displacement change caused by the inclusion remains poorly estimated.

Evaluating many candidate inclusions under new loads also raises a
computational question. Graph networks and neural operators have been used
to model mesh deformation and heterogeneous mechanics
\citep{pfaff,koja,goswami}, and video-based methods fit deformable models from
visual observations~\citep{jiang,physreal}. Structural reanalysis offers a
way to reuse the homogeneous solution when stiffness changes locally
\citep{kirsch1972,li2022,kirsch1991,wureanalysis}. We use elastic compliance to
define how forces within an inclusion interact and fit a short polynomial
filter to approximate the resulting displacement update. We compare the
filter with converged local reanalysis under the same mechanical reference
and prior to test how much the approximation changes the final forecast.

\section{The intervention and the mechanical benchmark}
\label{sec:benchmark}
\begin{figure}[htbp]\centering\fig{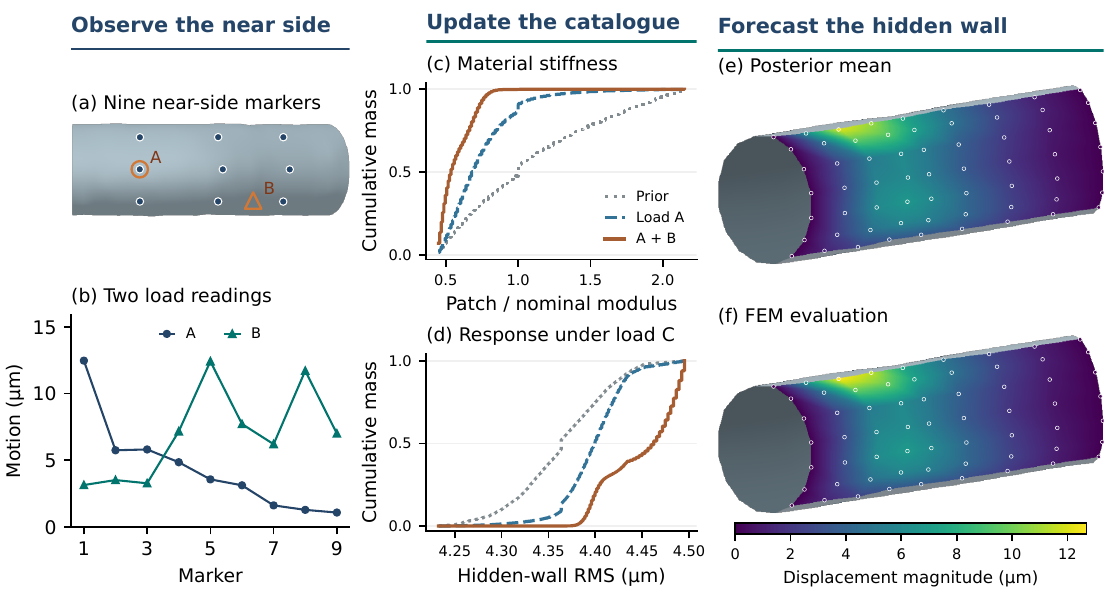}
\caption{From observed motion to a blind forecast on one held-out wedge tube. (a,b) Nine near-side markers supply three-component readings under loads A and B. (c,d) Cumulative catalogue probabilities before acquisition, after A and after both readings, for the modulus ratio and the area-weighted root-mean-square (RMS) displacement under a new load C. The final median modulus ratio is 0.54 (central 95\% catalogue interval 0.45--0.79). (e,f) The posterior mean vector field and independent finite-element target share a magnitude scale. Colour interpolates the scored markers for display. Relative vector error across all 63 markers is 2.92\%.}\label{fig:design}\end{figure}

For a rest geometry $G$, material $m$ and prescribed force vector $f_j$, the reference displacement $u_j$ solves the static small-strain problem $K(G,m)u_j=f_j$ on free degrees of freedom. In the wedge benchmark, inference receives the boundary conditions, geometry, nominal material, force catalogue and noise scale. The later tetrahedral studies treat support and other reference properties as uncertain. It observes
\begin{equation}
 y_j=H u_j+\epsilon_j,\qquad \epsilon_j\sim{\cal N}(0,\sigma^2 I_{27}),
 \quad j\in\{0,s\},
 \label{eq:obs}
\end{equation}
where $H$ reads three displacement components at each of nine near-side markers (Fig.~\ref{fig:design}). After load 0, a fixed or adaptive rule selects one of six second footprints. The targets are the opposite wall under four future loads and the modulus ratio $r=E_{\mathrm{patch}}/E_0$. Each future load has a component outside the span of its two acquisition loads, at least 0.816 of its norm across admissible triples.

The Synthetic Tubular Intervention Benchmark v1 contains 96 seeded LumenMorph geometries from four families (one straight, three curved), each split 12/4/8 for training, validation and testing \citep{geometry}. Scaling to published rubber-tube fixtures \citep{cam,gregory} gives mounted lengths of 16.69--28.16\,mm, inner radii of 3.15--3.40\,mm and nominal wall thicknesses of 0.19 or 0.38\,mm. Each tube has a prescribed stress-free shape, clamped end rings, $E_0=1$\,MPa and Poisson ratio 0.45.

\begin{figure}[htbp]\centering\fig{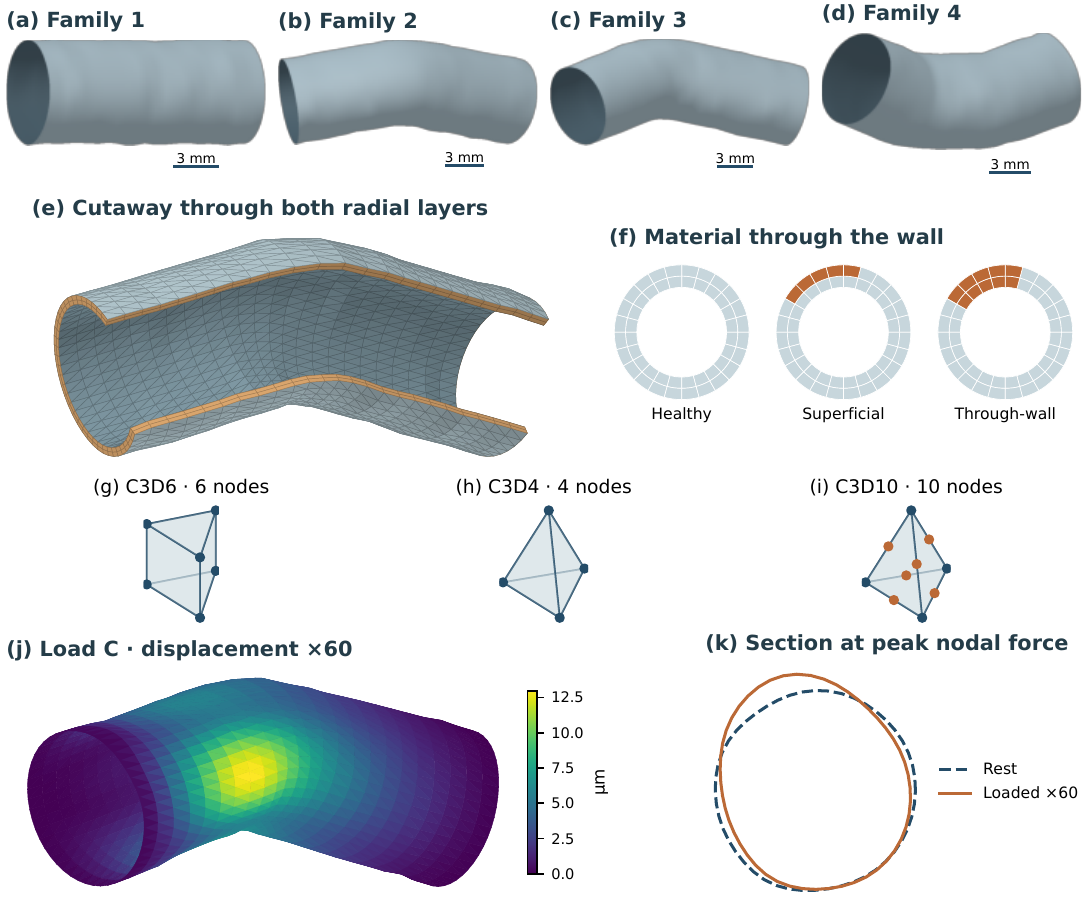}
\caption{One test geometry per family (first held-out replica), each with a 3\,mm scale bar. Removing one angular quarter of the family-2 wedge mesh exposes its lumen and both radial layers. The material sections show the three possible depth configurations schematically, with enlarged thickness. Below, nodal diagrams distinguish the six-node wedge used here from the linear and quadratic tetrahedra used for transfer and refinement. Copper points mark its six additional edge nodes. Bottom, full FEM under load C on the soft family-2 specimen. Colour gives mean nodal displacement magnitude per outer face. The rendered deformation and section use a 60-fold displacement magnification.}\label{fig:anatomy}\end{figure}

Figure~\ref{fig:anatomy} shows how the benchmark resolves curvature and inclusion depth. Each mesh has 24 axial cells, 48 circumferential cells, two radial layers and 4,608 six-node wedges (C3D6). Every geometry supplies a homogeneous, soft and stiff specimen. Patch location and extent vary off catalogue, affecting the outer layer or the full wall. Soft ratios span $[0.48,0.82]$ and stiff ratios $[1.28,2.10]$. Eleven radial footprints (summed nodal force magnitudes 1--3\,mN) yield $96\times3\times11=3,168$ finite-element (FEM) fields. All element Jacobians and loaded strain energies were positive. The maximum relative free-force residual was $3.57\times10^{-11}$ and maximum displacement 0.101\,mm.

A common $9\times16$ surface grid supplies nine visible and 63 area-weighted hidden markers, including hidden end-ring nodes. Independent noise per coordinate and acquisition has $\sigma=\max(0.0002\,\mathrm{mm},0.05\,\mathrm{RMS}_{0})$, where $\mathrm{RMS}_{0}$ is the nominal first visible response. Appendix~\ref{sec:rig} replaces these metric readings with pixels and a force readout.

\section{Estimators and evaluation}
\label{sec:methods}
We compute visible readings and hidden-wall displacements for candidate materials under the prescribed loads. The homogeneous mechanical reference is the displacement of the tube with uniform nominal stiffness for a given geometry, support and load. The inclusion contributes the difference between the displacements with and without that inclusion under the same conditions. Comparing predicted visible readings with the observations gives the weights used to average the candidate hidden-wall displacements under the new load.

\begin{figure}[htbp]\centering\fig{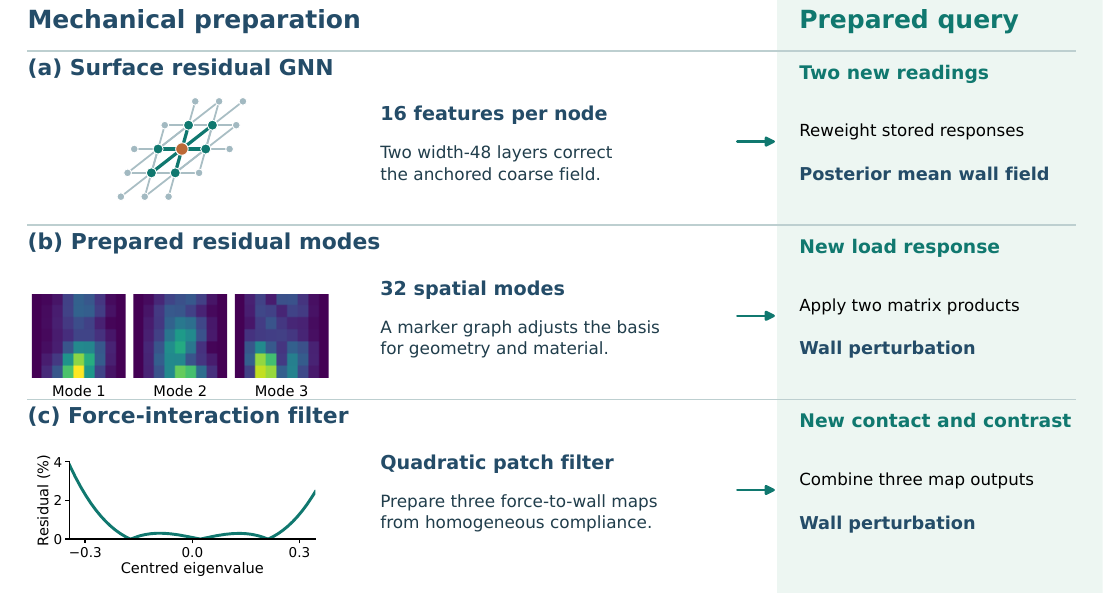}
\caption{What each operator prepares and what changes at query time. (a) Two updates reach the neighbours (teal) and their neighbours (grey) around a surface node (copper). (b) The first three of 32 residual modes are shown before graph adjustment. Colour gives vector magnitude on the unwrapped hidden wall, on one scale. (c) The curve gives the scalar inverse residual $|1-(1-x)p(x)|$ for the fitted stiff, through-wall branch over its theoretical interval. The three stored maps carry the response to the wall.}\label{fig:architectures}\end{figure}

Three operators serve different purposes (Fig.~\ref{fig:architectures}, Section~\ref{sec:materialupdate}). The surface network corrects a prediction built from a fine homogeneous field and a coarse inclusion response. The residual-mode operator prepares a reduced map for new loads. The patch filter uses the homogeneous mechanics to approximate the displacement change from a specified inclusion.

\paragraph{Forward response and catalogue.}
The unlearned baseline adds the coarse perturbation $\Delta u_c(m)=u_c(m)-u_c(1)$ to the fine homogeneous field $u_{0,h}$ at common nodes. All candidates reuse the fine nominal factorization. To correct this sum in local axial, circumferential and radial coordinates, the surface graph encodes position, response and candidate patch in sixteen features. Two width-48 mean-neighbour residual layers with hyperbolic tangents predict the vector correction. Controls are a parameter-matched nodal model (10,275 parameters), ridge regression on 21 features including neighbour differences, and the unlearned sum. Training used eight development geometries and seven loads (Adam, $10^{-3}$, batches of 32, 60 epochs). Four other geometries ranked ten-epoch checkpoints by perturbation error before v1 testing. Only v1 training data informed the later depth coefficients.

At eight candidate sites, fine or surrogate endpoint perturbations $S_-,S_+$ at $r=0.5,2$ define a one-pole approximation $ab/(1+\lambda a)$, $a=r-1$. Visible components under seven loads fit $0\leq\lambda\leq1$. Averaging the two endpoint estimates gives $b=[(\lambda-2)S_-+(1+\lambda)S_+]/2$. Even fine FEM endpoints require interpolation, and true patches can differ from catalogue sites and shapes.

\paragraph{Inversion and design.}
The prior assigns $1/17$ to healthy material and spreads $16/17$ uniformly over eight sites and 81 log-spaced ratios from 0.45 to 2.15. Equation~\eqref{eq:obs} gives posterior weights for the mean hidden field, median ratio and maximum posterior site mass. A fixed policy uses development-selected load 4. The active policy ranks six second loads by expected dispersion of the four future fields \citep{zhong}, using 32 stratified posterior draws and antithetic noise probes. Each policy sees only its two acquired responses.

\paragraph{Field error and evaluation protocol.}
For hidden reference vectors $u_i$ and area weights $w_i$ the relative error is
\begin{equation}
 e_u=\left(\frac{\sum_iw_i\|\hat u_i-u_i\|^2}
                         {\sum_iw_i\|u_i\|^2}\right)^{1/2}.
\end{equation}
We apply this metric to full displacements or to material perturbations $u(m)-u(1)$. Secondary metrics are vector RMSE (mm) and median $|\hat r/r-1|$. Each specimen contributes four future loads. Paired intervals use 10,000 geometry resamples, with patched curved tubes the primary comparison. We sealed blind predictions before scoring. Only the evaluator used hidden fields and labels to construct rival pairs. Prespecified targets were a 20\% active-acquisition error reduction with an interval excluding zero, a 10\% graph advantage over the nodal model in forward error, and twice the speed of the fastest adequate classical control. Failed controls count toward their requested denominators.

\subsection{A force-interaction graph for local stiffness changes}\label{sec:materialupdate}
Can the homogeneous tube predict a material change without a coarse heterogeneous solve? Write $K=K_0+aPDP^\top$, where $a=r-1$, $D$ is the nominal stiffness of affected elements in patch coordinates and $P$ embeds their free degrees of freedom. A force $f$ produces the homogeneous solution $u_0=K_0^{-1}f$ and patch force $q=DP^\top u_0$. Let $C=P^\top K_0^{-1}P$ be patch compliance, $T=DC$ its interaction operator and $J=H_wK_0^{-1}P$ the map to the hidden wall read by $H_w$. The exact wall change is $-aJ(I+aT)^{-1}q$. Each new patch requires restricting the homogeneous compliance.

\begin{figure}[htbp]\centering\fig{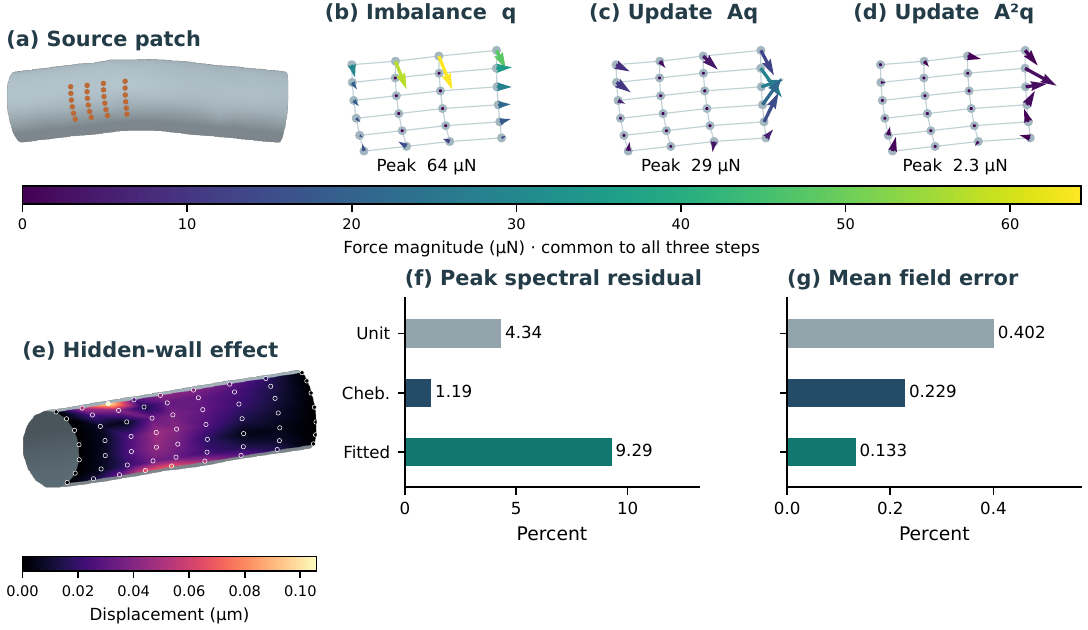}
\caption{A force update on a held-out tube. (a) Copper marks the outer nodes of the through-wall patch. (b--d) For the first future contact, arrows project the nodal forces $q,Aq,A^2q$ into the same plane, with $A=-b(T-I/2)$. Arrow lengths are normalized per panel for direction visibility. Colour uses one absolute force scale throughout. (e) The fitted sum projects to the opposite wall. (f,g) Worst sampled inverse residual on twelve tube--resolution combinations and mean displacement perturbation error on 48 tubes favour different filters.}\label{fig:propagation}\end{figure}

We replace the local inverse with a polynomial graph filter \citep{defferrard}. Each node carries three force components. The block $T_{ij}$ maps force at node $j$ to imbalance at node $i$ through the elastic tube. With $b=a/(1+a/2)$, the prediction is
\begin{equation}
 \widehat{\Delta u}_w=-bJ\sum_{k=0}^2\theta_k[-b(T-I/2)]^k q .
\end{equation}
Twelve development tubes supplied three fitted coefficients per depth and contrast sign (twelve scalars), minimizing area-weighted squared vector error divided by reference perturbation energy. The pointwise control uses the same imbalance, coefficient count and readout with only the diagonal $3\times3$ blocks of $T$. The fixed graph retains all blocks with unit coefficients. Figure~\ref{fig:propagation} follows one patch through the fitted filter.

\paragraph{Why centre the filter at one half?}
Assume $K_0\succ0$ on the free degrees of freedom and $0\preceq PDP^\top\preceq K_0$, as holds for a subset of the nominal elastic elements. Since $P$ selects free coordinates, $C\succ0$. Writing $A=K_0^{-1/2}P$, the assembly bound gives $ADA^\top\preceq I$. Its nonzero eigenvalues equal those of $D^{1/2}CD^{1/2}$ and hence of $S=C^{1/2}DC^{1/2}$. Thus $0\preceq S\preceq I$ and $T=C^{-1/2}SC^{1/2}$ has spectrum in $[0,1]$. In these symmetric coordinates, put $X=-b(S-I/2)$. For any $r>0$,
\begin{equation}
 a(I+aS)^{-1}=b(I-X)^{-1},\qquad \|X\|_2\leq\rho=|r-1|/(r+1)<1 .
\end{equation}
With unit coefficients the filter uses $I+X+X^2$. Its relative inverse residual is $X^3$ and the norm of its inverse-series remainder is at most $\rho^3/(1-\rho)$. Multiplication by $|b|\,\|JC^{-1/2}\|_2\,\|C^{1/2}q\|_2$ gives an absolute wall-error bound. The projected perturbation can be small, so its relative error requires its own denominator. For fitted coefficients let $p_\theta(x)=\theta_0+\theta_1x+\theta_2x^2$ and $R_\theta(x)=1-(1-x)p_\theta(x)$. Since $X$ is symmetric, diagonalization gives
\begin{equation}
 \|(I-X)^{-1}-p_\theta(X)\|_2 \leq
 \frac{M_\theta(\rho)}{1-\rho},\qquad
 M_\theta(\rho)=\max_{|x|\leq\rho}|R_\theta(x)|.
\end{equation}
Evaluating the endpoints and interior stationary points of the cubic bounds every admissible patch spectrum over the contrast intervals in Table~\ref{tab:cert}, a subset of the inverse catalogue. The same projection factor bounds the absolute wall error.

\begin{table}[tbp]\centering\small
\caption{Fitted coefficients and uniform relative inverse-residual bounds. Soft contrasts span $[0.48,0.80]$ and stiff contrasts $[1.30,2.05]$. Bounds use full-precision coefficients and round upward.}\label{tab:cert}
\begin{tabular}{lrrrr}\toprule
Inclusion & $\theta_0$ & $\theta_1$ & $\theta_2$ & $M_\theta$ (\%)\\\midrule
Through-wall, soft &1.000326&1.030154&0.990292&3.71\\
Through-wall, stiff &0.999161&1.038804&1.104041&3.83\\
Superficial, soft &1.001074&1.006105&0.815552&5.61\\
Superficial, stiff &0.997679&1.003910&1.358213&9.30\\\bottomrule
\end{tabular}\end{table}

\section{Hidden-field forecasts and material estimates}
The initial wedge benchmark tests whether two visible readings support both a hidden-wall forecast and a stiffness estimate (Table~\ref{tab:main}).
\begin{table}[tbp]\centering
\caption{Two-load predictions on 24 held-out curved geometries, each with two heterogeneous specimens and four future loads (192 requests per row). The ratio error is computed once per material case.}\label{tab:main}
\begin{tabular}{llrr}\toprule
Forward operator & Second force & Mean hidden error & Median ratio error\\\midrule
Anchored, unlearned & Fixed & 1.61\% & 17.95\%\\
Ridge correction & Fixed & 1.65\% & 16.90\%\\
Graph correction & Fixed & 1.62\% & 19.20\%\\
Graph correction & Active & 1.57\% & 17.48\%\\\bottomrule
\end{tabular}\end{table}
On curved patches the fixed graph's 1.62\% field error corresponded to $0.105\,\mu$m mean RMSE (95\% interval for relative error 1.31--1.95\%). Field error was 1.70\% on 16 patches from eight straight tubes. Curved-tube ratio errors were 15.34\% through-wall and 21.66\% superficial. Active selection missed its target (active minus fixed, $-0.047$ percentage points, interval $-0.251$ to $+0.135$). On eight earlier exposed bent geometries at 5\% noise, exact discrete-catalogue FEM gave 0.270\%, 0.187\%, 0.166\% and 0.158\% hidden error for random, calibrated fixed, material-focused and field-focused actions. The best improved 15.28\% over fixed.

On twelve further single-layer geometries, moving and resizing full-thickness patches away from catalogue sites raised graph ratio error from 8.45\% to 11.79\% and field error from 0.98\% to 1.15\%. Site accuracy fell from 71.88\% to 64.84\%, conditional 95\% ratio coverage from 95.05\% to 88.28\%. Centred finite differences under four future loads gave mean log-modulus derivative errors of 25.34\% for the graph, 24.45\% for ridge and 12.92\% for fine one-pole FEM.

\paragraph{Which computation improves the field?}
On a separate one-layer cohort (20 calibration, 24 test tubes), matching the graph's nominal mesh to the $24\times48$ reference instead of $16\times32$ cut field error from 6.60\% to 1.11\% (5.49 points, 95\% interval 4.91--6.08) and ratio error from 60.62\% to 11.77\%. Matched-reference ridge, nodal and unlearned errors were 1.09\%, 1.11\% and 1.20\%. In an earlier known-material comparison, errors were 20.90\% for coarse FEM, 1.01\% anchored, 0.61\% ridge and 0.66\% for both graph and nodal networks. A branch/trunk operator reached 0.88\% on those exposed geometries. Nominal resolution accounted for most of the gain.

We calibrated the maximum normalized field error over three materials and four loads separately from the maximum absolute log-ratio error over two patches, within each geometry. Under exchangeability the 19th of 20 ordered scores gives each endpoint a marginal 90\% split-conformal bound \citep{lei}. The matched-anchor field radius was 3.52\% of the predicted norm (target 5\%). The ratio interval required a factor of 1.63 (target 1.25). Each covered 23/24 held-out geometries.

\section{Forecasting with an uncertain mechanical reference}\label{sec:reference}

This experiment tests whether visible readings improve forecasts when the mechanical reference itself is uncertain. We withheld resolution ($n=32$ or 40 axial cells, $2n$ circumferential), thickness (uniform multiplier 0.94--1.06) and distal fixation (clamped or partly released at unknown phase). Given nominal geometry, parameter ranges and frontal stereo readings under two known loads, we forecast three-component displacements at 63 hidden markers under 16 further contacts for healthy, superficial and through-wall specimens. Training, selection, calibration and prospective test cohorts were separate and balanced across three curved families. Paired 95\% intervals use 10,000 geometry resamples.

\paragraph{Joint inference over reference and material.}
Eight homogeneous FEM endpoints spanning resolution, thickness and support yield 28 references by interpolation at seven thicknesses. A coarse library spans eight patch sites and two depths. Soft and stiff endpoints define one-pole responses at 21 log-spaced contrasts. Each reference scales the material perturbation by its response norm relative to the catalogue. For candidate reference $a$, material $m$, predicted visible pixels $h_{am}$, hidden field $u_{am}(f)$ and noise covariance $\Sigma$, the estimator returns
\begin{equation}
  \widehat u(f)=\sum_{a,m}p_{am}(y)\,u_{am}(f),\qquad
  p_{am}(y)\propto\pi_{am}\exp\!\left[-\tfrac12\|y-h_{am}\|_{\Sigma^{-1}}^2\right].
\end{equation}
The prior weights references and the 17 material branches uniformly (healthy plus 16 site/depth pairs), then weights contrasts uniformly within branch. Fit and calibration each used 24 geometries. The conformal multiplier is the 23rd ordered geometry maximum across specimens and forces. Errors use the true weighted RMS norm and radii use the predicted norm.

\begin{table}[tbp]
\centering\small
\caption{Unknown-reference forecasts on 36 held-out geometries. Errors are relative weighted RMS percentages. The first column averages 1,152 patched-specimen requests, the remaining columns use all 1,728 requests. The fixed mean reference was selected on development geometries.}\label{tab:references}
\begin{tabular}{lrrr}\toprule
Method & Patched mean & Overall mean & Overall P95\\\midrule
Nominal FEM, $n=24$ & 12.44 & 12.39 & 23.36\\
Nominal FEM, $n=32$ & 5.31 & 4.98 & 10.72\\
Fixed mean reference & 5.07 & 4.67 & 10.89\\
Adaptive reference, homogeneous material & 3.74 & 2.99 & 7.93\\
Adaptive reference and material, $n=16$ & 3.43 & 2.86 & 7.06\\
Adaptive reference and material, $n=24$ & 3.40 & 2.85 & 7.13\\
Adaptive reference and material, $n=32$ & 3.40 & 2.86 & 7.16\\\bottomrule
\end{tabular}
\end{table}

On 36 test tubes, adaptive $n=24$ reduced patched error by 36.0\% versus nominal $n=32$ FEM (95\% interval 24.9--45.1\%) and 32.9\% versus the best fixed reference (22.2--41.8\%, Table~\ref{tab:references}). Mean absolute whole-field RMS error was $6.72\times10^{-5}$\,mm. A constant 16.07\% radius covered all requests on 36/36 tubes. A radius based on six mechanical predictions averaged 12.26\% and covered 34/36 tubes. Preparing its three catalogues and endpoints took a median 77.73\,s. Only 13/1,728 simultaneous radii were at most 5\%. A marginal band calibrated for one uniformly selected request covered 34/36 preselected requests. Of its 113 radii at most 5\% across all 1,728 fields, three had errors above 5\%.

\paragraph{Cost of the alternative explanations.}
On six preselected tubes, exhaustive FEM required $33\times8=264$ factorizations per tube. With the same observations, priors and averaging, patched error was 2.970\% for exhaustive FEM and 3.015\% and 3.076\% for $n=24,16$ transfer. Median Mac setup plus inference took 333.87, 27.19 and 16.24\,s, respectively. Over 36 tubes, $n=16$ cost 40\% as much as adaptive $n=32$ for a $+0.028$-point error difference ($-0.059$ to $+0.116$). On 720 observation pairs from 24 other tubes, precomputing quadratic likelihood and field bases cut median/P95 update times for 16 fields to 0.140/0.163\,ms from 0.506/0.564\,ms on the Mac, and to 0.537/3.506 from 4.691/5.851\,ms on a concurrently loaded Linux workstation. Outputs agreed within $2\times10^{-15}$\,mm. Timing excludes acquisition and registration.

\subsection{Changing the element changes the mechanical reference}\label{sec:mesh}
We tested twelve geometries each with nominal dimensions, intermediate thickness, extreme slenderness or localized stenosis. On nominal tubes the frozen wedge-trained graph gave 4.93\% perturbation error against wedges (C3D6) and 41.07\% against linear tetrahedra (C3D4). Its original band covered every required field on only 1/48 tubes across strata.

The 48 exposed tubes fitted a rank-32 residual to the tetrahedral perturbation. Inference recomputed coarse candidate fields, the fine homogeneous field and eight reference solutions with tetrahedra. After calibration on 24 separate tubes we tested 48 new geometries, twelve per stratum. Mean perturbation error fell from 37.53\% (coarse) to 16.05\% (frozen graph with consistent inputs) and 12.58\% (adapted). Total-field error against this tetrahedral reference was 0.49\% with known material and 2.80\% from two visible readings. The band covered all twelve fields on 44/48 tubes (exact 95\% interval 80.0--97.7\%), including 9/12 at extreme lengths. Only 26/96 inclusions had median stiffness estimates within a factor of 1.25.

With two radial layers held fixed on four representative tubes, refining quadratic tetrahedra (C3D10) from 24 to 32 rings changed total displacement by up to 4.79\% and the perturbation by 5.57\%. A two-resolution Richardson extrapolation selected there gave 5.15\% perturbation error on twelve reserved designs (two inclusions, four forces), versus 8.40\% for coarse quadratic FEM and 35.54\% for the frozen wedge graph. Changing elements also changes surface-load quadrature. In 45 neo-Hookean equilibria on three tubes, departures from load superposition reached 4.47\% at the 20\% nominal strain target and linearization error reached 13.08\%. Across the three tubes, linearization error was 0.21--0.41\% at 0.50\% measured strain and 1.94--3.89\% at 4.61--4.88\%. The hidden-field errors below measure agreement with a specified discrete small-strain model.

\begin{figure}[htbp]\centering\fig{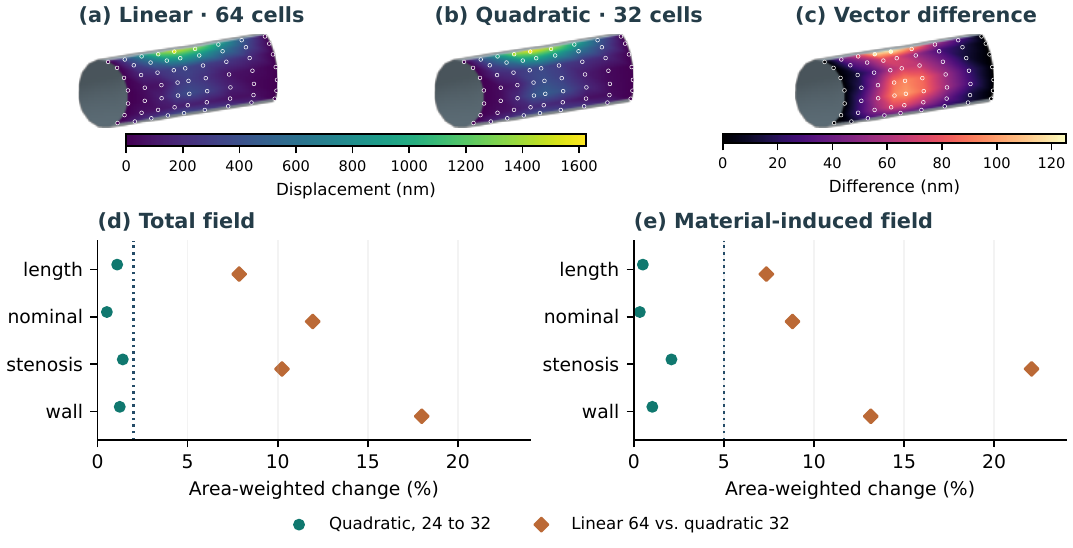}
\caption{Reference sensitivity under matched traction. (a,b) Altered-state wall displacement on the nominal tube at the same 411,264 free displacement components, on a shared scale. (c) Magnitude of the vector difference at the same markers, with its own scale. (d,e) On four tubes, circles compare the last two quadratic refinements and diamonds compare linear and quadratic elements at equal displacement count. Changes are relative to the finer quadratic field; dotted lines mark the 2\% and 5\% tolerances.}\label{fig:meshsensitivity}\end{figure}

Figure~\ref{fig:meshsensitivity} compares the last quadratic refinement with the cross-order discrepancy. Supplement S1 reports replica 00 of curved family 2 in each test stratum, as selected by protocol. Both ends were clamped, a superficial patch had twice the nominal modulus, and a Gaussian traction acted on the exterior faces. Meshes had $n$ axial, $2n$ circumferential and $n/4$ radial cells, with $n=8,16,32,48,64$ for linear tetrahedra and $n=8,16,24,32$ for quadratic tetrahedra. The last quadratic refinement changed hidden displacement by 0.52--1.40\% and its material perturbation by 0.33--2.09\%. At 411,264 free displacement components each, the finest linear and quadratic solutions differed by 7.86--17.99\% and 7.36--22.08\%, respectively, relative to the quadratic field. No tube met the prespecified criteria across refinements and element orders (2\% for total displacement and compliance, 5\% for perturbation). Blind errors therefore retain their discrete references.

\subsection{How much information do the two readings add?}
To measure the contribution of each reading, we retrospectively replayed the uncertain-reference C3D4 cohort and the known-fixation patch-filter cohort with zero, one or two readings (48 tubes each, Table~\ref{tab:cohorts}). Within each cohort we held candidate libraries, forces, priors and reference uncertainty fixed. Two readings reproduced every sealed field within $2\times10^{-14}$\,mm absolute plus $3\times10^{-10}$ relative tolerance. The zero-reading controls were the prior mean and the nominal healthy field. With material supplied, the tetrahedral control gave 0.49\% total error over 6,144 catalogue fields on the nominal 32-ring mesh. On the known-fixation bank, supplying the true material for the same 96 blind inclusions gave 0.0030\% over 384 requests. Both blind tasks withhold material identity. Only the tetrahedral task also withholds the reference. The known-fixation catalogue includes the true material.

The second reading improved total-field error by 1.17 percentage points on tetrahedra (paired 95\% interval 0.83--1.53) and 0.34 points on the patch bank (0.23--0.45). On the latter, the first reading gained just 0.03 points ($-0.07$ to $+0.15$). The nominal healthy field alone gave 5.87\% and 1.45\% total error, respectively. On tetrahedra the mean absolute RMS errors for zero, one and two readings were 0.118, 0.083 and 0.058\,$\mu$m.

The paired element-order diagnostic tests the same linear-reference estimator with quadratic observations and targets. It reuses 12 tubes from the uncertain-reference C3D4 cohort, one per stratum and curved family, with inference libraries and noise draws held fixed. With two readings, total error rose from 3.10\% against C3D4 to 41.29\% against C3D10. On C3D10 the zero- and one-reading errors were 49.58\% and 41.98\%. The second reading gained 1.87 points on the paired linear references (95\% interval 0.98--2.68) and 0.69 points on C3D10 (0.41--1.03), improving 11/12 quadratic forecasts at the tube level. The two-reading material-effect error reached 281.8\%, and simultaneous coverage fell from 11/12 to 0/12. On four prespecified tubes both additional refinement checks met every load-specific tolerance and surface refinement changed the gain by at most 0.021 points (Supplement S1, Tables S5--S6 and Fig.~S2).

\begin{figure}[tbp]\centering\fig{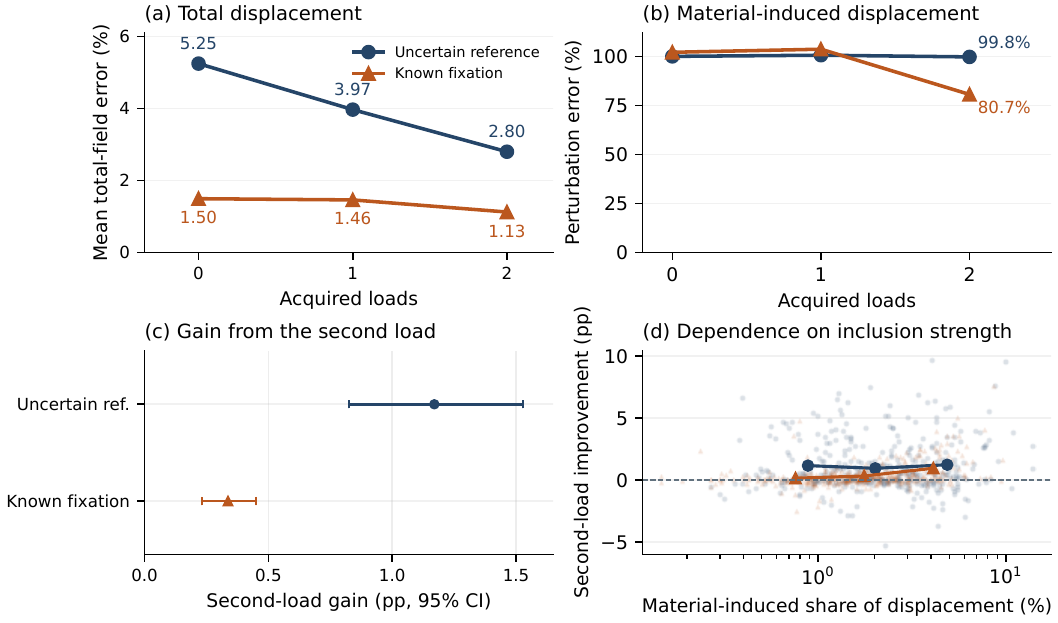}
\caption{What does a second reading add? (a,b) Mean vector error from zero to two readings on two exposed cohorts, for the total field (576 requests per cohort) and its material-induced component (384 heterogeneous requests). (c) Paired 95\% intervals for the second-load gain in total-field accuracy, in percentage points (pp). (d) Field-level gains on patched specimens against the material-induced share $100\,\|u_m-u_h\|/\|u_m\|$, with matched healthy field $u_h$. Lines connect means in three calibration-defined bins per cohort.}\label{fig:budget}\end{figure}

\begin{table}[tbp]\centering\small
\caption{Mean hidden-wall error (\%) as observations arrive. Each cohort has 48 tubes. Total error uses 576 fields. Perturbation error uses the 384 fields with inclusions, relative to their matched healthy fields. Zero readings means averaging the original prior.}\label{tab:budgets}
\begin{tabular}{lrrr}\toprule
Cohort and target & Zero & One & Two\\\midrule
Uncertain reference, total &5.25&3.97&2.80\\
Uncertain reference, perturbation &100.1&100.7&99.8\\
Known fixation, total &1.50&1.46&1.13\\
Known fixation, perturbation &102.2&103.7&80.7\\\bottomrule
\end{tabular}\end{table}
Predicting no material effect gives 100\% perturbation error on each altered field (Table~\ref{tab:budgets}). The tetrahedral estimate subtracts a posterior-weighted healthy response, the patch filter the known healthy field. The reference uses the true load and support. From the lowest to highest calibration-defined third of material strength, two-reading total error rose from 1.93\% to 4.70\% on tetrahedra and from 0.76\% to 2.49\% on the patch bank (Fig.~\ref{fig:budget}).

\begin{figure}[htbp]\centering\fig{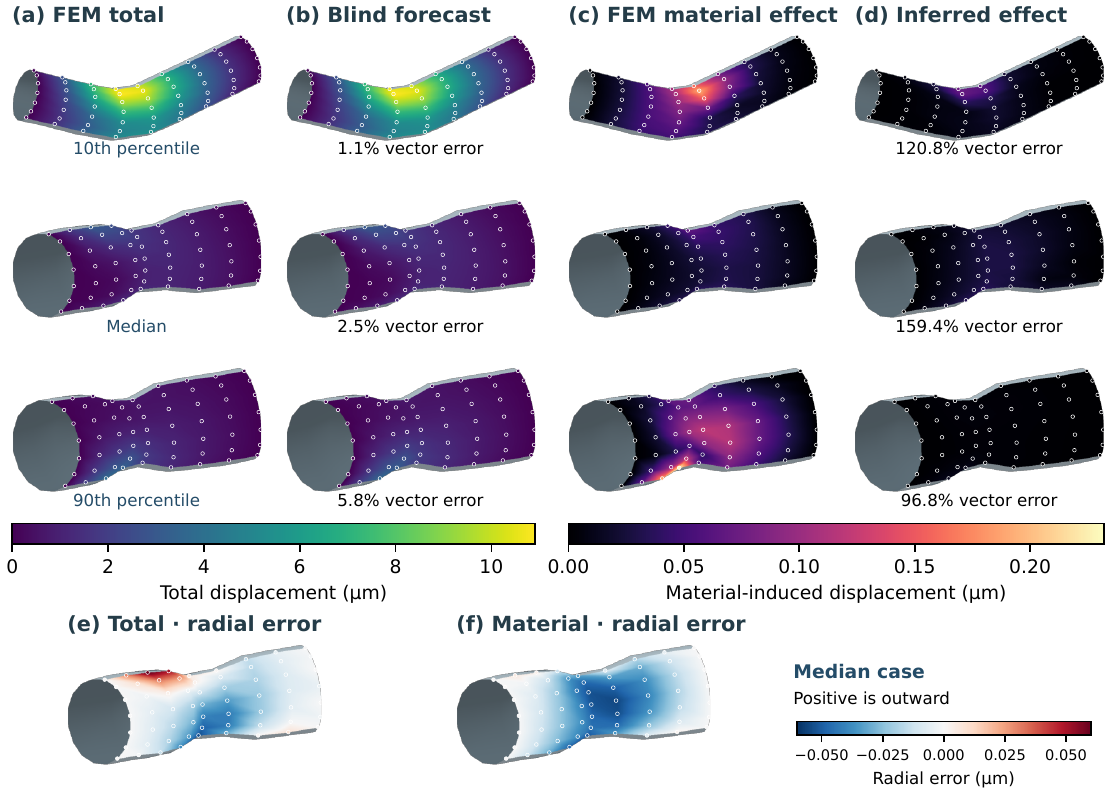}
\caption{Blind tetrahedral predictions. Rows in (a--d) select the nearest ranks to the 10th, 50th and 90th percentiles of total-field error among 384 heterogeneous two-reading requests. Total displacements and material-induced changes have separate magnitude scales, each shared across tubes and methods. Annotated errors use area-weighted three-component vectors at all 63 hidden markers. (e,f) For the same median case, the signed radial component of prediction minus FEM uses the rest-ring centroid to define the outward direction. Colour interpolates marker values for display.}\label{fig:wallfields}\end{figure}

The spatial comparison in Fig.~\ref{fig:wallfields} separates error in the total displacement from error in the smaller material-induced component on the same contacts.

\subsection{Does the patch filter change a blind forecast with uncertain mechanics?}
On the same clamped 32-ring tetrahedral model, we retrospectively compared a numerically converged material update with fixed, Chebyshev and fitted quadratics. Each computed stereo displacements and four future fields at 21 ratios, eight sites and two depths, using the same measurements, prior and transfer to 28 references. Each operator supplied its own predicted readings to the likelihood and therefore its own posterior weights. The fitted update retained its twelve wedge-trained coefficients. The earlier 2.80\% forecast used a rank-corrected field and one-pole interpolation. We calibrated each band on the same 24 tubes and tested it on the same 48 others.

All four posteriors gave 2.77\% blind total-field error (Table~\ref{tab:matched}). The fitted-minus-reference difference was $-0.00010$ percentage points (paired 95\% interval $-0.00039$ to $+0.00017$). Both had 34.6\% median absolute modulus-ratio error. With material and nominal mesh supplied, the wedge-trained filter gave 0.272\% perturbation error across 6,144 tetrahedral endpoint fields. Its advantage over the other quadratics appeared in this conditional calculation.

\begin{table}[tbp]\centering\small
\caption{Material updates under reference uncertainty on 48 tetrahedral tubes. Total error covers 576 fields. Blind perturbation error uses the 384 altered fields and a posterior-weighted healthy baseline. Conditional error uses the known nominal reference and endpoint material. Coverage requires all twelve fields per tube inside its calibrated band.}\label{tab:matched}
\begin{tabular}{lrrrrr}\toprule
 & Blind total & Blind perturb. & Conditional perturb. & Covered & Median radius\\
Update & (\%) & (\%) & (\%) & (of 48) & (\% predicted RMS)\\\midrule
Converged series & 2.7709 & 96.86 & $<10^{-7}$ & 42 & 7.09\\
Fixed quadratic & 2.7705 & 96.78 & 0.846 & 41 & 7.03\\
Chebyshev quadratic & 2.7714 & 96.93 & 0.548 & 42 & 7.08\\
Fitted quadratic & 2.7708 & 96.83 & 0.272 & 42 & 7.07\\\bottomrule
\end{tabular}\end{table}
The reference summed the centred inverse series through degree 34 for six force vectors. The relative local-equation residual was below $6\times10^{-15}$ and the maximum relative endpoint discrepancy from independent heterogeneous FEM was $1.4\times10^{-9}$. Given eight archived healthy anchors, median CPU preparation took 5.42\,s per quadratic catalogue and 44.74\,s for the converged series. Posterior inference for three specimens took 118--122\,ms per tube across methods. Section~\ref{sec:repeated} times prepared GPU queries at fixed and changing contrasts.

\section{Which materials can the visible measurements distinguish?}
\paragraph{Prospective depth experiment.}
Using 96 labelled patches on 48 v1 training tubes, we fitted a multiplier $q_d$ per depth to minimize scattering error across seven loads. On 24 fresh curved geometries, loads 0 and 4 probed one through-wall, one superficial and one healthy specimen each. We compared full-only and two-depth catalogues. The graph uses $r_{\rm eff}=1+q_d(r-1)$. FEM computes depth-specific endpoints with the same ratio interpolation.

\begin{table}[htbp]\centering
\caption{Blind depth experiment on 24 additional geometries. Material errors are medians on 24 inclusions per depth. The mean future field error pools all 48 patches. Depth calls are conditional on the specimen containing a patch.}\label{tab:depth}
\begin{tabular}{lrrrr}\toprule
&\multicolumn{2}{c}{Ratio error (\%)}& Field error & Correct depth\\
Operator and catalogue & Through & Superficial & (\%) & (of 48)\\\midrule
Graph, calibrated full only & 8.12 & 28.20 & 1.56 & ---\\
Graph, both depths & 3.57 & 25.52 & 1.49 & 33\\
FEM, full only & 8.27 & 28.42 & 1.37 & ---\\
FEM, both depths & 7.10 & 21.66 & 1.27 & 36\\\bottomrule
\end{tabular}\end{table}
Adding depth reduced superficial ratio error by 9.51\% for the graph (95\% interval $-6.52$ to $41.46$\%) and 23.78\% for FEM ($-8.54$ to $58.96$\%). Neither cleared the 20\% target with a positive lower bound. Among known inclusions, depth accuracy reached 68.75\% (58.33--79.17\%) and 75.00\% (64.58--85.42\%), respectively (Table~\ref{tab:depth}). Both maximum-posterior-site rules also called a patch in all 24 healthy controls.

\paragraph{Local criterion for a predictable target.}
Let $J_V$ and $J_W$ map a small parameter change to visible readings and future wall displacement. A direction $v$ is invisible to first order if $J_Vv=0$. All such directions preserve the target exactly to first order if and only if $\ker J_V\subseteq\ker J_W$. With noise, even a weakly visible direction matters if its wall response is large.

We evaluated 16 site/depth branches at healthy contrast on each of 48 nominal clamped 32-ring tetrahedral tubes. A unit parameter change represents one modulus-ratio unit or a 6\% thickness change. Let $V$ be the pixel Jacobian divided by the acquisition noise standard deviation and $W$ the wall Jacobian with area weights and equal weight across the four future loads. Centred differences (dimensionless step $10^{-4}$) matched analytic contrast derivatives within $3\times10^{-8}$ relative error. Across 768 branches the median weak and strong visible singular values were 1.27 and 5.83 noise units per unit parameter change. The corresponding median wall sensitivities were 0.089 and 0.171\,$\mu$m per unit. Each $V=U\Sigma Q^\top$ had full column rank, giving worst local gain $\sup_{v\ne0}\|Wv\|_2/\|Vv\|_2=\|WQ\Sigma^{-1}\|_2$. Its median was 0.080\,$\mu$m per noise unit (interquartile range 0.041--0.159). Across 48 tubes the median local catalogue gain over branches correlated weakly with mean absolute blind error on the eight patched requests (Spearman $r_s=0.18$).

\paragraph{Constructing alternatives.}
The graph fitted $q_{\rm through}=1.30$ (upper search bound) and $q_{\rm superficial}=0.64$. At a fixed site, equal $q_d(r-1)$ gives identical graph fields under every load, leaving depth preferences to the prior and contrast range. To test whether the mechanics also admit rivals, the evaluator kept each true patch footprint, switched its depth and searched 81 contrasts against the two noiseless visible responses. A fresh two-layer FEM solve verified the selected rival. Only this evaluator knew the true footprint.

Let $\mu,\mu'$ be the two fixed candidates' noiseless readings (54 coordinates). With equal priors and the Gaussian noise in Eq.~\eqref{eq:obs}, projection along their difference gives optimal binary accuracy
\begin{equation}
 P_{\rm correct}=\Phi(d/2),\qquad d=\|\mu-\mu'\|_2/\sigma .
 \label{eq:bound}
\end{equation}
At $d\leq2\Phi^{-1}(0.90)\simeq2.5631$, even an observer knowing both fields has at most 90\% binary accuracy. This held for 43/48 exact pairs, including all 24 superficial and 19/24 through-wall targets. Their median modulus differences were 23.63\% and 28.21\%, respectively, while their future hidden fields differed by just 0.20\% and 0.43\%. The bound applies to these constructed pairs under the specified noise model.

\begin{figure}[htbp]\centering\fig{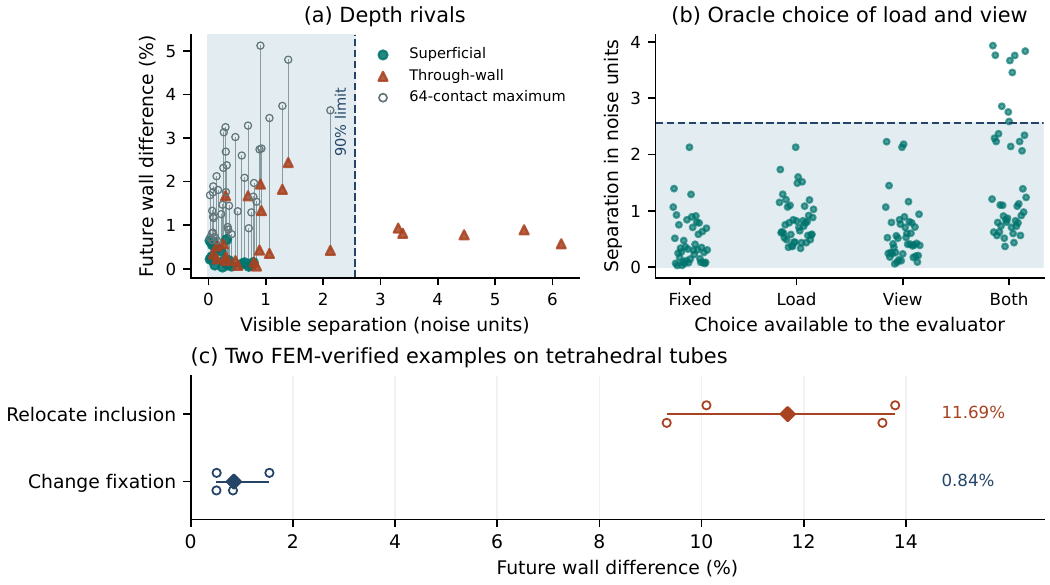}
\caption{Can similar readings conceal different futures? (a) Among 48 depth rivals, filled symbols give the mean wall difference over four future loads. For 43 pairs below the discrimination threshold (shaded, $d\leq2.5631$), open symbols show the maximum across 64 further contacts. (b) An oracle can separate nine of those pairs by choosing both load and view. (c) Two selected tetrahedral pairs fall below the same threshold. Changing fixation ($d=0.075$) or relocating an inclusion ($d=0.735$) gives different wall discrepancies. Circles show the four forces and diamonds their means.}\label{fig:pairs}\end{figure}
Among six loads and four views, an oracle knowing both materials separated 0/43 pairs by load alone, 0/43 by view alone and 9/43 by joint choice (Fig.~\ref{fig:pairs}). Appendix~\ref{app:acquisition} tests policies that choose before the second reading.

We applied the same threshold to stereo readings on the 48 tetrahedral tubes (72 image coordinates across two loads, $\sigma=\sqrt{2}\,0.02$\,px). Errors are independent across coordinates and acquisitions. A shared noisy rest image would require whitening by the joint covariance. In the catalogue, 3,600 of 5,376 equal-depth location pairs and all 1,536 clamped versus partly released fixture pairs met the threshold. For each group we re-solved by FEM the pair with the largest predicted wall discrepancy. Relocation gave $d=0.735$ and 11.69\% mean relative wall difference across four future loads (79\,nm mean RMS). Changing the fixture gave $d=0.075$ and 0.84\% (6\,nm). These extremes illustrate possible discrepancies. Their frequency requires verification of more pairs.

\begin{figure}[htbp]\centering\fig{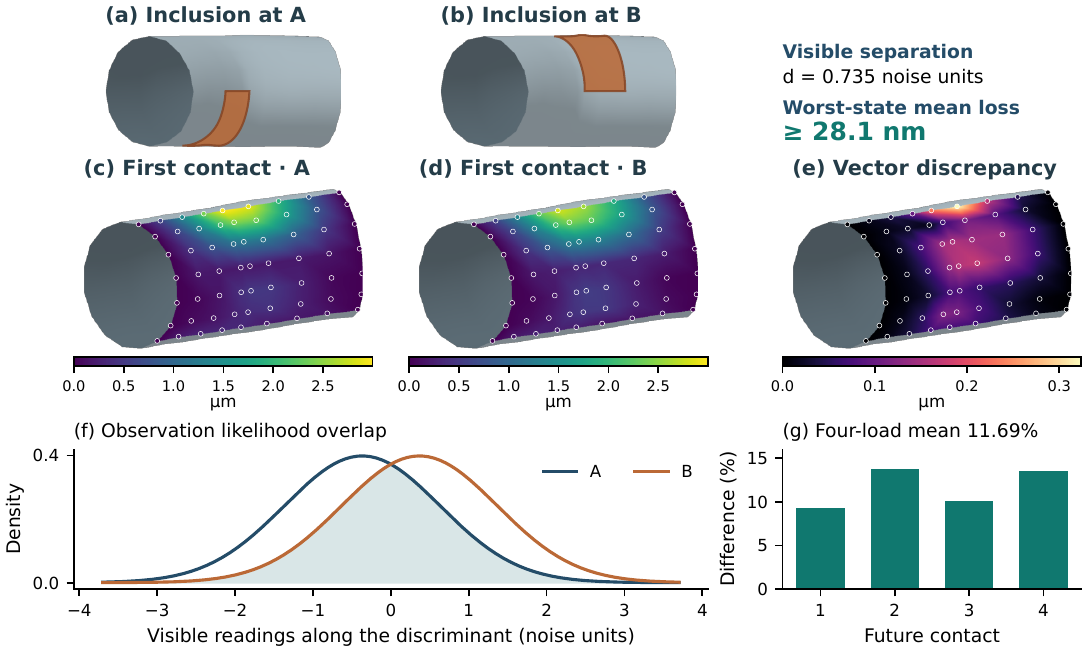}
\caption{The same through-wall inclusion ($E/E_0=2$) at two locations on one tetrahedral tube. The footprints are drawn through the rest shell. Under the first future contact, the two hidden-marker fields share a scale. Their vector difference has a separate scale. Shading between markers is illustrative. The two Gaussian observation densities overlap along the optimal discriminant ($d=0.735$). Across four future contacts, the mean relative wall difference is 11.69\%. Equation~\eqref{eq:pairrisk} bounds the worst-state expected absolute loss for this selected pair from below by 28.1\,nm.}\label{fig:relocation}\end{figure}

Figure~\ref{fig:relocation} shows where the relocation changes the hidden field under the first future contact.

These FEM pairs also bound the error of any forecast from the two readings. Let $L$ be the mean of the four area-weighted RMS norms and $\Delta=L(u_1-u_2)$ the absolute separation of two target fields. For observation densities $p_1,p_2$, the triangle inequality gives $L(\hat u-u_1)+L(\hat u-u_2)\geq\Delta$ at every observation. Integrating against $\min(p_1,p_2)/2$ bounds the larger expected loss from below by $\Delta\int\min(p_1,p_2)/2$, the two-point argument of Yu~\citep{yu1997}. For the equal-covariance Gaussian readings in Eq.~\eqref{eq:bound}, this yields
\begin{equation}
 \inf_{\hat u}\max_{i=1,2}\mathrm{E}_i L(\hat u-u_i)\ \geq\ \Delta\,\Phi(-d/2).\label{eq:pairrisk}
\end{equation}
Using unrounded FEM separations gives 28.1\,nm for relocation and 2.85\,nm for fixture change. The relocation bound is about 4\% of either state's mean four-load hidden displacement (0.704\,$\mu$m RMS). For scale, the separate stereo experiment gave 2.32\,$\mu$m marker RMSE at 0.02\,px noise (Appendix~\ref{sec:rig}), a different error measure. If these two states are the only possibilities and both responses are known, choosing the nearer visible mean attains the bound. Under either state it incurs loss $\Delta$ with probability $\Phi(-d/2)$.

\paragraph{Which material information survives the instrument?}
The median worst-contact difference across the 43 eligible pairs was 1.76\%. For nonnegative mixtures of these 64 forces with total coefficient at most one, linearity and the triangle inequality bound the absolute difference by the largest absolute dictionary response for that pair. One rival pair differed by 5.12\% under a later frontal contact (RMS $4.25\times10^{-5}$\,mm). None of 64 extra contacts resolved it at the visible threshold. On the 36-geometry cohort, joint inference reduced local-ratio MAE across 72 inclusions from 0.513 (development median) to 0.409, a 20.2\% gain (9.4--30.4\%), but simultaneous intervals excluded healthy stiffness only once. On 24 further tubes with two fixed contacts and 0.20 inter-frame error correlation, sixteen exposures per contact reduced MAE from 0.463 to 0.357 at 0.02\,px (23.0\% gain, 9.1--36.6\%) and yielded 0.285 at 0.005\,px. At 0.02\,px, calibrated three-class sets made ten definite altered-stiffness calls, including one healthy specimen. Soft, healthy and stiff coverage was 21/24, 23/24 and 23/24. Bursts assumed a stationary specimen and contact.

\section{Preparing the mechanics for repeated queries}\label{sec:repeated}
We separate the cost of preparing the mechanical response from the cost of querying it at new loads and contrasts.

\subsection{What the force-interaction graph adds}
\begin{figure}[!htbp]\centering\fig{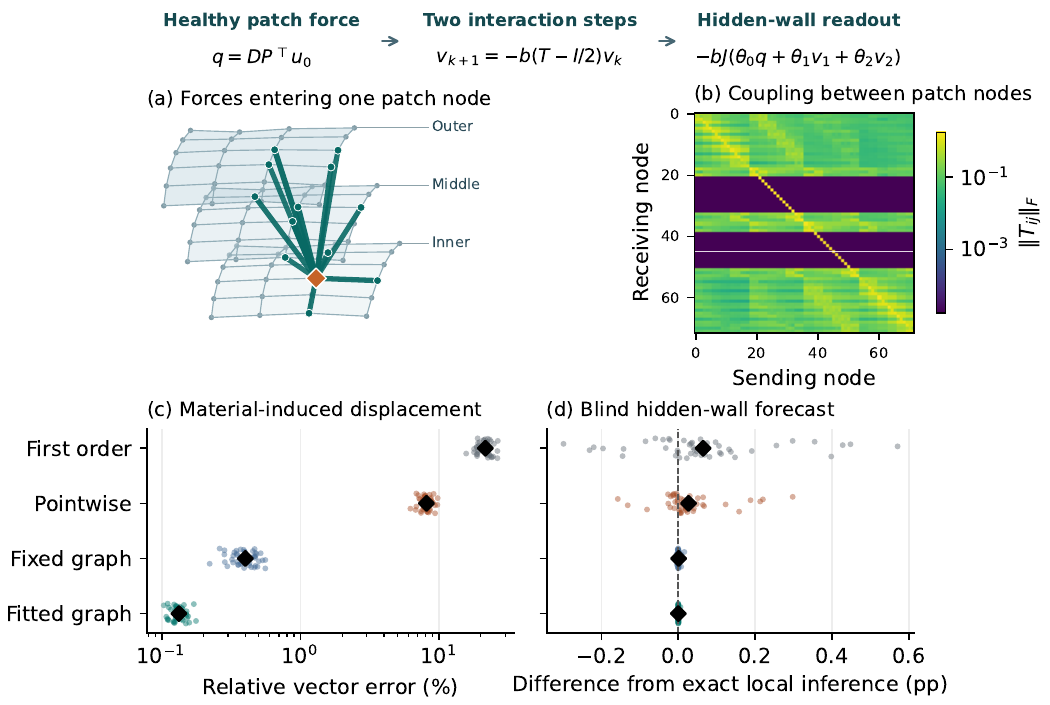}
\caption{From a contact force to a hidden-wall update. Starting at $v_0=q$, two propagation steps mix three-component nodal forces before projection by $J$. (a) Twelve strongest interactions into one receiving node (diamond) of a through-wall patch. The three radial node layers are separated for display. Line width encodes the norm of each interaction block. (b) Norms of all $3\times3$ blocks of $T$ on the same patch, clipped five decades below the maximum. White marks the receiving row. (c,d) Perturbation error and blind forecast difference from exact local inference given the same two readings. Dots are held-out tubes and diamonds are means.}\label{fig:materialupdate}\end{figure}

We calibrated on 24 tubes and tested 48 more across three curved families. Eight patch footprints per tube varied in centre and width, each with two depths and two contrasts per depth. Independent direct solves on the same 16-ring mesh supplied the reference for four new contacts. Figure~\ref{fig:materialupdate} relates the interaction blocks to the per-tube errors. Across 6,144 requests, mean perturbation errors were 21.55\% (first order), 8.115\% (pointwise), 0.402\% (fixed graph) and 0.133\% (fitted graph). The paired reduction versus pointwise was 98.36\% (family-stratified 95\% interval 98.32--98.40\%). A degree-two Chebyshev interpolant gave 0.229\%. In blind inference, two noisy nine-marker readings per specimen and known fixation gave the graph and exact inverse identical mean total-field error (1.126\%) and median ratio error (12.06\% on 96 inclusions). Specimens belonged to the candidate bank. The graph band covered all twelve fields on 47/48 tubes.

Including setup and 1,000 contacts on six tubes, median CPU times for a fixed 32-material bank were 2.585\,s for the graph, 2.582\,s for the exact local inverse and 9.545\,s for repeated global factorization with an output adjoint. Changing all 32 ratios at each contact on fixed patches took 2.792\,s for the graph and 2.939\,s for an exact inverse reusing patch eigensystems. A cubic Chebyshev filter reduced perturbation error from 0.134\% to 0.025\% and took 3.053\,s.

Against 32-ring references on six development tubes, the same graph coefficients gave 33.35\% perturbation error with 16-ring mechanics and 0.310\% with matching 32-ring mechanics. Transolver groups spatial samples into learned states \citep{transolver}. For the load-conditioned comparison we used its original catalogue and compressed both prepared force maps by their singular spectra. On six tubes from the 48-tube primary cohort, held out from timing development, perturbation errors were 0.316\% for the fine-mesh graph and 6.645\% for Transolver. Each GPU call received four host contact descriptors and returned 128 wall fields to host memory. Across 192 paired blocks, median times were 0.167 versus 1.913\,ms (11.47-fold), P95 times 0.194 versus 2.292\,ms (11.84-fold). Prior CPU preparation took 33.1\,s median for the graph's fine homogeneous maps versus 27.8\,s for Transolver's mechanical inputs.

\paragraph{Direct comparison with exact local updates on the GPU.}
On the same six tubes we varied resolution (16 or 32 rings), material count (one or 32) and whether contrast changed at each contact. Each GPU call took one host contact descriptor and returned the healthy and all requested heterogeneous fields. Both methods shared traction, homogeneous factorization, patch restriction and single-precision arithmetic with reduced-precision tensor cores disabled. At fixed ratios, warm timings selected dense or low-rank maps (relative Frobenius tolerance $10^{-5}$). Changing ratios used full polynomial bases or local eigensystems. For $n_f$ traction coordinates and $n_w$ wall outputs, the quadratic stores three $n_w\times n_f$ maps per site and depth. Each contact needs three matrix--vector products shared across contrasts. A rank-$k$ exact spectral update stores $k(n_f+n_w+1)$ scalars, projects the input once, then weights and projects $k$ modes per contrast. We checked outputs against direct heterogeneous FEM before timing. Direct local solves for individual fixed ratios matched the spectral maps within $2\times10^{-6}$ relative norm and reused their query timings.

\begin{figure}[!htbp]\centering\fig{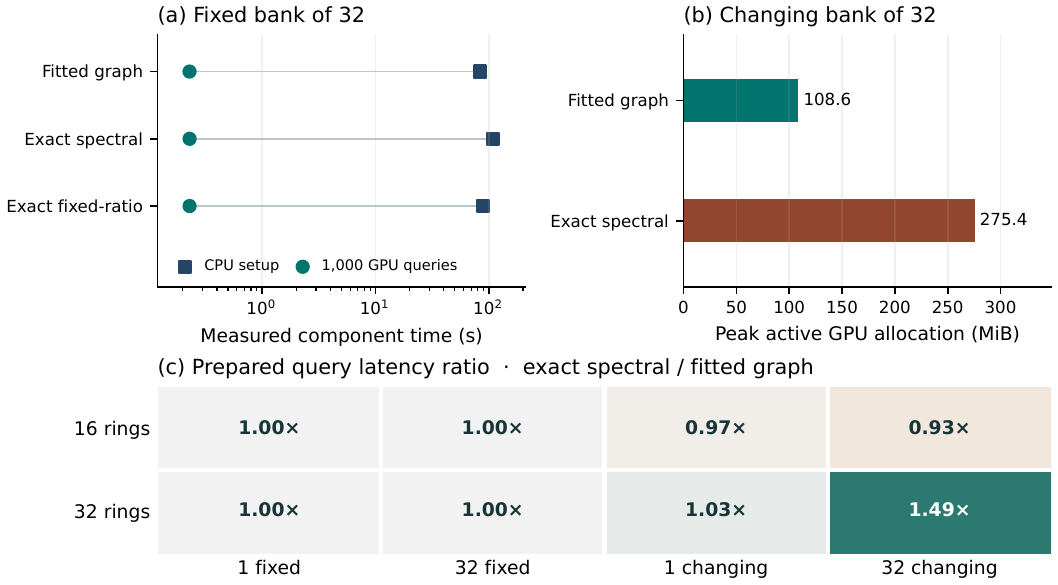}
\caption{Cost for 32 materials on six 32-ring tubes. (a) Median CPU preparation time and median summed time for 1,000 GPU contacts at fixed contrasts, plotted separately on a logarithmic axis. The fixed-ratio control reuses the verified exact-map query time. (b) Peak active GPU allocation with changing contrasts. (c) Median prepared latency of exact spectral inversion divided by that of the fitted graph across eight workloads. Above one favours the graph. At 32 rings and 32 changing contrasts the two queries took 0.705 and 0.474\,ms.}\label{fig:exactcost}\end{figure}

At 32 rings and 32 fixed materials, graph and exact maps both took 0.236\,ms per contact (median of six tube medians). Including CPU setup, device upload, warm qualification and 1,000 sequential queries, elapsed component sums were 83.29\,s for the graph, 89.51\,s for the fixed-ratio exact solve and 108.61\,s for the exact spectral inverse (Fig.~\ref{fig:exactcost}). With 32 contrasts changing at each contact, graph and spectral medians were 0.474 and 0.705\,ms, P95 times 0.491 and 0.725\,ms, and peak active CUDA allocations 108.6 and 275.4\,MiB. Relative perturbation error against the spectral update averaged 0.157\% on those changing contacts. At 16 rings the spectral query was faster (0.265 versus 0.285\,ms). With one changing material at 32 rings the graph and exact queries took 0.266 versus 0.273\,ms. The advantage depends on both mesh and bank size.

Shared compliance preparation per new footprint took a median 0.35\,s at 16 rings (405 local degrees of freedom) and 9.50\,s at 32 rings (1,377). At 32 rings, both depths added 0.53\,s for polynomial terms or 3.52\,s for spectral factors. Six geometries ran concurrently on the Intel Core Ultra 9 275HX. Isolated GPU timings used an NVIDIA GeForce RTX 5070 Ti Laptop GPU (driver 580.178.04, PyTorch 2.9.0, CUDA 12.8) and exclude file I/O, training and acquisition. At 32 rings with 32 materials, median total costs kept the same ordering at every measured prefix through 1,000 contacts. Fixed contrasts favour the more accurate exact map at essentially equal query latency. Changing many contrasts favours the fitted filter.

On the same twelve tube--resolution combinations, we evaluated the relative scalar inverse residual $|1-(1-x)p(x)|$, $x=-b(\lambda-1/2)$, at the computed spectra including zero and 65 evenly spaced ratios in each soft and stiff range. Maximum residuals were 4.34\% for unit coefficients, 9.29\% for the fitted quadratic and 1.19\% for a degree-two Chebyshev interpolant. The fitted polynomial improves the area-weighted displacement endpoint it was trained on, while Chebyshev gives the smallest worst sampled spectral residual.

\section{Qualifying the mechanical reference}
The independent confirmation tests whether rebuilding the mechanical reference restores accuracy on new tubes. The paired diagnostic in Section~\ref{sec:reference} motivates this change. Across those twelve tubes, the linear-to-quadratic change in the homogeneous wall field was 49.70\% of total C3D10 displacement. The corresponding change in the material perturbation was 0.95\%. In the two-reading linear-inference forecast the homogeneous residual alone was 43.52\% and the perturbation residual was 5.60\% on the same scale. These are vector contributions. Their cross term affects the combined error. At identical mesh, load and support, the four order changes were 44.19--69.69\%. Subdividing the wall radially changed C3D4 by 13.01--31.68\% and C3D10 by 0.09--0.14\%. Figure S3 and Supplement S1 document the traction, reaction, support and geometry controls.

We rebuilt the prior response library using homogeneous C3D10 solutions on nominal tubes at three thicknesses and two distal supports, interpolated to seven thicknesses per support. Each factorization supplies the six acquisition and forecast load responses. We transport the material perturbation from a nominal 24-ring mesh to these anchors using separate visible and hidden norm ratios. The alternatives reuse C3D4 perturbations or compute the patch update on the quadratic stiffness. The fitted graph keeps the original coefficients. Fixed polynomials, a three-vector Galerkin solve and a converged Chebyshev series share its mechanics (Supplement S1). The estimator receives nominal geometry, the specified loads and noisy optical readings. The catalogue marginalizes thickness and support. Global compliance is scaled jointly across observing and forecasting loads.

We used the same 12 new tubes to select the 32-ring C3D10 references with global compliance uncertainty and to calibrate the bands. The 24 confirmation tubes were independent of those 12 tubes and the earlier paired diagnostic. They use 40- or 48-ring C3D10 meshes, continuously integrated tractions, out-of-catalogue material patches and independently perturbed radii, modulus, thickness and support. Acquisition and hidden evaluation use the same solved specimen. We set a practical gain of 0.25 percentage points per reading before confirmation and required the second-reading interval to exclude zero. The band targets were simultaneous coverage of at least 20/24 tubes and a mean diameter below 0.50\,$\mu$m.

On the independent cohort the two-reading forecast errs by 0.091\,$\mu$m. The first and second readings reduce error by 2.00 [1.09, 2.88] and 0.95 [0.62, 1.33] percentage points (mean and paired 95\% interval). Relative to matched C3D4, the two-reading error fell by 94.5\% [93.5, 95.4] (Table~\ref{tab:qualifiedreference}). With the same anchors and prior, converged reanalysis and the frozen filter gave 2.8456\% and 2.8450\% total error. The difference was less than 0.001 percentage points. Each computed its posterior from the same readings.

\begin{table}[tbp]\centering\small
\caption{Independent quadratic qualification on 24 tubes. Mean relative vector RMS error uses all 288 future fields for total displacement and the 192 altered fields for the material-induced component. Coverage requires all twelve future fields per tube to lie in the bands. Diameter is mean calibrated full width in the area-weighted norm.}\label{tab:qualifiedreference}
\setlength{\tabcolsep}{5pt}
\begin{tabular}{lrrrrrr}\toprule
Inference & Zero & One & Two & Material & Covered & Diam. ($\mu$m) \\ \midrule
C3D4 reanalysis & 60.78\% & 57.05\% & 56.61\% & 278.9\% & 21/24 & 11268.523\\
Homogeneous C3D10 & 6.19\% & 4.29\% & 3.51\% & 100.0\% & 24/24 & 19.990\\
C3D10 plus transferred C3D4 perturbations & 6.19\% & 4.24\% & 3.37\% & 102.4\% & 24/24 & 12.165\\
C3D10 with frozen graph & 6.18\% & 4.24\% & 3.39\% & 113.1\% & 24/24 & 2.285\\
24-ring C3D10 with converged reanalysis & 6.18\% & 4.24\% & 3.39\% & 113.1\% & 24/24 & 2.294\\
Selected C3D10 & 5.79\% & 3.79\% & 2.84\% & 103.4\% & 22/24 & 0.806\\
\bottomrule\end{tabular}\end{table}

\section{What carries over to measurements?}
To test sensitivity to image noise and load variability, we simulated opposing stereo pairs that independently supplied observations and third-load readouts. Across 864 cycles per noise level, increasing image noise from 0.02 to 0.10\,px raised curved-patch field error from 2.56\% to 2.69\% and ratio error from 26.17\% to 39.84\%. On 48 curved patched specimens (144 repeats per level), true third-load RMS displacement averaged 5.21\,$\mu$m. Force variability contributed 0.077\,$\mu$m mean RMS. Mean prediction errors were 0.081 and 0.087\,$\mu$m against nominal force, 0.128 and 0.136\,$\mu$m against realized force. These RMS components need not add. Stereo readout over all evaluable markers gave 2.32 and 11.59\,$\mu$m RMSE. Adaptive force selection failed to improve graph forecasts on 24 further tubes (Appendix~\ref{sec:rig}).

Eleven physical rubber-tube configurations test transfer to pressure-driven collapse. Given one-sided observations, fitted compliance predicted the next eight hidden markers with 1.164\,mm RMSE over 368 requests. Persisting the last observed oval shape gave 1.055\,mm. Appendix~\ref{app:transfer} examines this failure and a separate silicone-phantom control.

\section{Discussion}
The homogeneous mechanical reference accounted for most of the forecast discrepancy in the paired element-order test. Replacing the frozen filter by the converged inverse changed the 41\% error by less than 0.002 percentage points. Rebuilding the homogeneous library recovered the largest part of the lost accuracy on the independent synthetic cohort.

Superficial and through-wall inclusions can produce nearly identical visible motion. If they also move the hidden wall similarly, averaging their forecasts can predict that wall accurately while stiffness remains ambiguous. At fixed Poisson ratio, scaling all moduli and forces together preserves displacement. Force calibration removes this scale ambiguity but leaves depth, thickness and support to resolve.

With the corrected mechanics the old band multiplier covered 19/24 independent tubes. We used the same 12 tubes to select the reference and calibrate the bands, which covered 22 of 24 tubes in the independent confirmation cohort at a mean diameter of 0.806\,$\mu$m, above the 0.50\,$\mu$m target. The normalized optical residual ranked hidden errors poorly (Spearman $\rho=0.08$).

The fastest operator depends on what changes between queries. Reduced FEM gave the fastest queries for known material. When material changes, graph interactions capture displacement changes that pointwise updates miss. The exact local inverse remains competitive for fixed materials. With 32 changing contrasts, spectral queries were faster at 16 rings, while the fitted graph used less time and memory at 32 rings. New measurements alone require only posterior reweighting. A change in distal support, however, raised hidden error to 22.25\% in a separate perturbation test.

On 58 public brackets, adding two coarse elastic fields and training new weights reduced graph error from 45.44\% to 9.67\%. Both fields already encode force transmission from pin to bolts. Pointwise correction reached 9.71\% on the same inputs (Appendix~\ref{sec:brackets}). On quadratic tubes, classical two-grid extrapolation outperformed the wedge-trained graph. Tubular intervals quantify variation across seeded geometries from shared generator families with single inclusions and a prescribed material law.

The virtual rig isolates the material contribution by loading the same simulated tube with and without an inclusion and reading the entire hidden wall. We quantify optical error separately with known calibration and marker identity. The measured pressure and indentation controls address separate physical prediction tasks. A matched physical test would measure three force footprints on a tube of known stiffness and fixture compliance, then compare the sealed third-load forecast with independent measurements of the opposite wall. Reconstructed scenes would also require calibrated forces and registered metric marker displacements.

\section{Conclusion}

In the simulated tubes, two visible displacement readings supported
accurate forecasts of the opposite wall under a new load, even when the
effect of a local stiffness change remained poorly estimated. The paired
element-order test traced most of the forecast discrepancy to the
predicted response of the homogeneous tube. On 24 independent simulated
tubes, the homogeneous quadratic model recovered most of the lost
accuracy, reducing mean relative hidden-wall error from 56.61\% to 3.51\%.
The full estimator reduced it to 2.84\%. Relative error in the displacement
change caused by the inclusion remained above 100\%.

The second reading met the prespecified criterion for a practically
useful improvement. The prediction bands covered 22 of 24 tubes, with a
mean width above the target. For repeated evaluations of candidate
inclusions, the compliance-based graph filter reused the homogeneous
factorization and achieved nearly the same blind forecast accuracy as
converged local reanalysis under matched conditions. Classical polynomial
and Galerkin approximations gave comparable forecast accuracy.

These results establish what can be forecast under the simulated loads
and observation conditions studied here. The measured pressure and
indentation controls address separate prediction tasks. A direct physical
test would compare a forecast based on two visible readings with an
independent measurement of the hidden wall under a third prescribed load.

\section*{Data, code and attribution}
The simulated tubular fields, observations, fitted coefficients, sealed predictions, timing records and scoring code are publicly available in version 1.2.0 of the reproduction dataset \citep{wallreproduction} (\href{https://doi.org/10.57967/hf/11868}{DOI 10.57967/hf/11868}, revision \href{https://huggingface.co/datasets/icirauqui/hidden-wall-deformation-reproducibility/tree/5aad8a5a1766ae21ea80faaefbe17e7f4428d6e6}{\texttt{5aad8a5a1766}}). The deposited archives include the frozen code, protocols, and scoring records for the primary comparisons, the paired element-order test, and the independent high-fidelity qualification. The corresponding author can provide additional review material for the virtual rig, brackets and figures. Transolver uses the MIT-licensed irregular-mesh implementation \citep{transolver}. Sources include the tubular geometry collection \citep{geometry}, Whalen, Beyene and Mueller's SimJEB v4 decks and fields \citep{simjebdata,simjeb}, Gregory, Agarwal and Lasenby's Cambridge marker data (CC BY-NC 4.0) \citep{cam} and the OpenCAS silicone phantom \citep{suwelack}.

\appendix
\counterwithin{figure}{section}
\counterwithin{table}{section}
\counterwithin{equation}{section}
\section{Which tubes support each main comparison?}
The mesh study reuses four tetrahedral test tubes, and GPU timings reuse tubes from the patch-filter experiment. Table~\ref{tab:cohorts} traces these overlaps.
\begin{table}[htbp]\centering\small
\caption{Cohorts and their reuse. Counts are tubes or bracket designs. Identifiers match the review-package manifests. Selection chooses a model or implementation. Calibration sets the band multiplier.}\label{tab:cohorts}
\begin{tabularx}{\linewidth}{@{}l>{\raggedright\arraybackslash}X>{\raggedright\arraybackslash}X@{}}\toprule
Comparison & Fitting, selection and calibration & Test and reuse\\\midrule
C3D4 reference, \texttt{test-*} & 48 prior tubes fit the residual model. The four patch updates reuse frozen coefficients. Both analyses calibrate bands on the same 24 \texttt{cal-*} tubes. & Same 48 tubes in four strata, 576 fields per arm. Four of these geometries supply the mesh check.\\
Paired element-order test & Same inference libraries and band multipliers as in the original C3D4 test. & 12 of the 48 C3D4 test tubes, 144 fields per reference. Four of the twelve supply the load-specific refinement check.\\
Independent quadratic qualification & 12 new tubes select the reference and calibrate bands. & 24 further tubes, 288 fields per arm. Both sets are distinct from the paired test.\\
Depth rivals, \texttt{depth-*} & 48 benchmark training tubes fit depth response. The separate detection threshold uses 16 validation tubes. & 24 tubes, 48 inclusions. The 64-force analysis uses the 43 qualifying pairs.\\
Patch filter, \texttt{update-test-*} & 12 fitting and 24 band-calibration tubes. & 48 tubes, 6,144 conditional and 576 blind fields.\\
GPU cost & Six timing-development tubes from the patch cohort. & Six other tubes from that cohort, replicas 02/03 per family, at both resolutions.\\
Residual modes, \texttt{speed-*} & 90 training, 24 selection, 24 separate \texttt{speedcal-*} band-calibration tubes. & 48 tubes, four reserved contacts each.\\
Reference adaptation, \texttt{adapt-*} & 24 design and 24 band-calibration tubes. & 36 tubes, 1,728 forecasts. Joint and marginal bands share requests.\\
SimJEB, \texttt{primary}/\texttt{authors} & 266/57 training/selection designs on the primary split and 252/52 on the designer split. & 58 and 77 test designs, respectively. Four loads per design.\\\bottomrule
\end{tabularx}\end{table}

\section{Choosing a second load or view}\label{app:acquisition}

\paragraph{Choosing a view from observations alone.}
On 24 new curved geometries, a policy ranked measurements by posterior linear correlation with depth. Conditional on a patch, it regressed depth on the 27 visible components with instrument noise included in their covariance. Joint selection improved depth accuracy from 30/48 to 33/48 for the graph and 34/48 to 38/48 for FEM. Gains of 6.25 points ($-6.25$ to $18.75$) and 8.33 ($-8.33$ to $25.00$) missed the ten-point target with a positive lower bound. The second view exposed some forecast markers, motivating a common hidden sector.

\paragraph{Forecasting a wall hidden from all candidate views.}
On 48 further curved geometries (96 patches, 48 healthy controls), all four views hid the same 54 scored markers. On the preceding 24 development geometries, the best of 24 fixed load--view pairs was load 3 at a half-turn shift for the graph and load 6 at a three-quarter turn for two-depth FEM. Active estimators saw identical first readings and sealed their choice before the second. Four future loads remained withheld.

Under the first-observation posterior, let $C$ be the covariance of catalogue scattering coefficients, $B_j$ the map to candidate measurement $j$, and $G$ the Gram matrix of the four hidden-field bases, area-weighted and normalized by each future load's homogeneous energy. The best linear update reduces expected weighted field variance by
\begin{equation}
 J_j=\mathrm{tr}\!\left[(B_j C B_j^\top+\sigma^2 I)^{-1}
                          B_j C G C B_j^\top\right].
\end{equation}
We rank actions by $J_j$ and forecast with the full Bayesian likelihood. Equal-budget controls use seeded random, field-focused load-only or view-only, and depth-focused actions.

\begin{figure}[htbp]\centering\fig{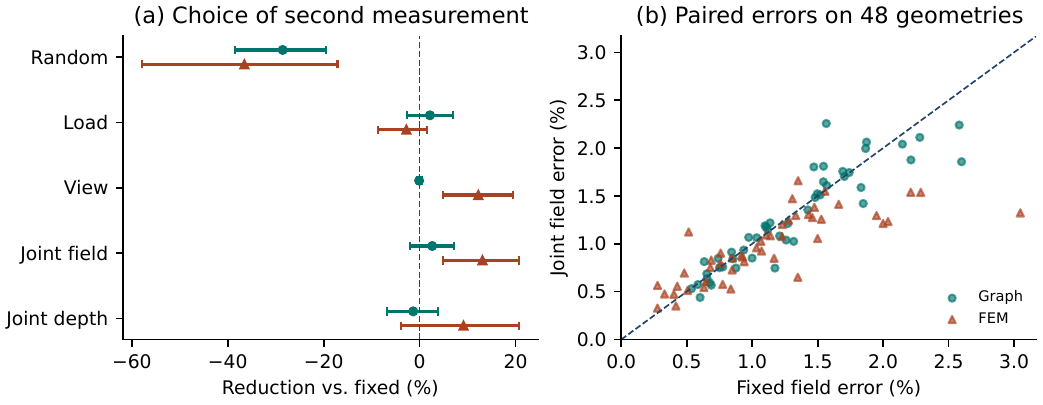}
\caption{Forecasts on the common hidden sector. Left, mean relative error reduction from each model's best development-fixed pair, with paired 95\% geometry intervals. Positive values favour the indicated policy. Right, each point averages two patched specimens and four future loads per test geometry. Points below the diagonal favour joint selection.}\label{fig:joint}\end{figure}
All policies returned 384/384 heterogeneous forecasts. Joint selection reduced graph error from 1.30\% to 1.27\% (2.62\% improvement, $-2.05$ to $7.18$\%), missing the primary 20\% target. For FEM, error fell from 1.12\% to 0.97\% (13.09\%, 4.82--20.62\%). View-only FEM reached 0.98\%, load-only 1.15\% (Fig.~\ref{fig:joint}). Random selection was worse than fixed for both. Joint selection changed median ratio error from 12.87\% to 16.22\% for the graph and 12.53\% to 11.75\% for FEM. The two joint policies falsely called patches in 34/48 and 30/48 healthy controls. The FEM endpoint was secondary.

\paragraph{When should the inverse report a material?}
With fixed loads 0 and 4 on 16 v1 validation geometries, we tested a healthy null using $z=(y-\mu_{\rm healthy})/\sigma$. Under the stipulated noise and a fixed healthy reference, $\|z\|^2$ is chi-squared with 54 degrees of freedom. Projection onto leading material-perturbation singular vectors (95\% squared singular-value mass) gives rank four or five. At a 5\% upper-tail cutoff, both matched-reference tests rejected 1/16 healthy specimens. The full test detected 1/16 superficial and 9/16 through-wall cases, the projected test 4/16 and 12/16. A coarser one-layer healthy reference produced 16/16 false alarms with either rule. Failure to reject leaves material unreported.

With fields and both readings fixed in the two observation-budget cohorts, we varied the healthy prior mass over $1/17,1/3,1/2,0.8,0.95$. A definite healthy or altered call required at least 0.95 posterior mass. On tetrahedra the original $1/17$ prior called 20/96 inclusions and 4/48 healthy specimens altered (120/144 abstentions). At $1/3$ only one inclusion was called altered, with zero healthy alarms and 143 abstentions. Across all five priors mean field error varied only from 2.78 to 2.80\%. On the in-catalogue bank, the original $1/3$ healthy mass (weight 16 versus one for each of 32 alternatives) gave 25/96 correct altered calls, zero healthy alarms and 119 abstentions. Both analyses are retrospective and use constructed populations.

\section{Three loads on a virtual test rig}\label{sec:rig}
\begin{figure}[htbp]\centering\fig{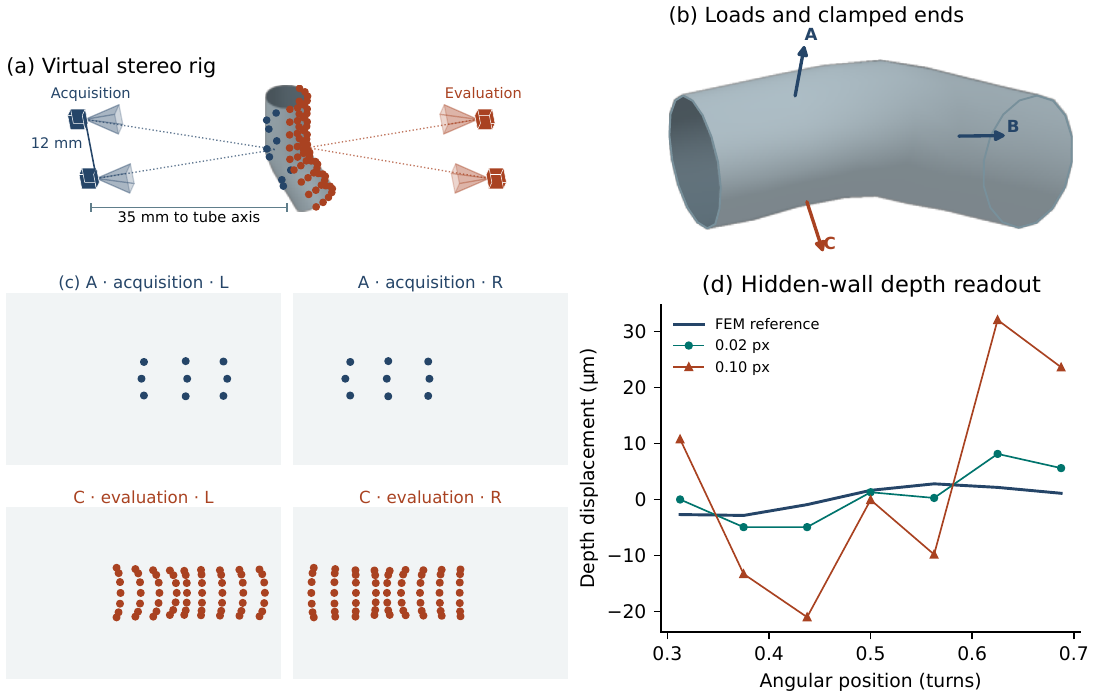}
\caption{Virtual stereo acquisition and independent readout. (a) Rectified camera pairs with 12\,mm baselines and 35\,mm stand-off; frusta indicate orientation. (b) Nominal resultant loads on the clamped tube. (c) Archived loaded marker coordinates in the two 1,920 by 1,200 pixel planes of each pair (first repetition, 0.02\,px noise). The acquisition pair observes load A and the opposite pair evaluates C. Only finite projections are shown. (d) Triangulated depth changes at seven mid-ring markers for load C under both noise levels, compared with the discrete FEM values.}\label{fig:rig}\end{figure}

The virtual rig used two opposing rectified stereo pairs (12\,mm baseline, 1600\,px focal length, 35\,mm stand-off, Fig.~\ref{fig:rig}). Ray tests required each labelled marker to remain visible in both cameras before and after loading. The two hypothetical image precisions (0.02 and 0.10\,px) shared standardized draws of independent Gaussian noise. Calibration and marker identities were known.

Each cycle applied three radial tractions of known footprint and direction, returning to rest between loads. Force amplitudes varied by 2\% standard deviation. Independent load-cell noise per component was 0.5\% of the nominal resultant magnitude. Measured forces and loaded-minus-unloaded pixels (variance $2\sigma^2$) informed the forecast under the nominal third force. Only the evaluator knew its realized amplitude, opposite-side pixels and full FEM field, enabling independent optical scoring in millimetres and mechanical scoring at all 63 hidden markers.

\paragraph{Fixed-load repeatability.}
Loads 0, 4 and 7 on 96 geometries, three materials and three repeats gave 864 cycles per noise level. The opposite pair triangulated 53,163/54,432 requested displacements at each level (vector RMSE 0.00232 and 0.01159\,mm). The main analysis uses the 48 patched specimens from 24 curved test tubes and averages repeats within specimen for ratio error.

\paragraph{Can choosing the second force help?}
On 24 further geometries (eight per curved family, two patched specimens and one healthy specimen each), the graph used load 0 to rank six second forces by linearized posterior variance reduction at load 7. The full inverse used both pixel readings. All four operators shared cameras, force budget, sites and full-wall material parametrization. Twelve separate calibration geometries selected each operator's best fixed force at each noise level. Policies sealed choices before the next reading and forecasts before scoring. At 0.02\,px the graph target was a 20\% improvement with a positive lower 95\% paired geometry-bootstrap bound and at least 90\% coverage.

\begin{table}[htbp]\centering
\caption{Mean relative hidden-field error (\%) over all 144 heterogeneous cycles on 24 additional curved tubes with two patched specimens each. Each fixed comparator was selected on development data for its operator.}\label{tab:rig}
\begin{tabular}{lrrrr}\toprule
& \multicolumn{2}{c}{0.02\,px} & \multicolumn{2}{c}{0.10\,px}\\
\cmidrule(lr){2-3}\cmidrule(lr){4-5}
Operator & Fixed & Active & Fixed & Active\\\midrule
Graph & 2.508 & 2.516 & 2.752 & 2.752\\
Ridge & 2.455 & 2.465 & 2.733 & 2.733\\
Unlearned anchor & 2.572 & 2.572 & 2.774 & 2.774\\
Fine FEM & 2.405 & 2.448 & 2.779 & 2.712\\\bottomrule
\end{tabular}\end{table}

At 0.02\,px the graph chose the fixed force in 137/144 patched cycles and improved by $-0.30$\% ($-0.87$ to $+0.20$\%). Random selection gave 2.618\% field error. At 0.10\,px all 144 choices matched fixed (Table~\ref{tab:rig}). Median ratio errors were 23.71\% fixed and 24.73\% active at lower noise and 46.12\% for both at higher noise. All policies returned 72/72 healthy forecasts per level. Fine FEM's secondary improvement at 0.10\,px was 2.39\% ($-0.65$ to $+5.34$\%).

\section{Comparisons on other operators and structures}
\subsection{Residual modes with prepared or load-conditioned evaluation}
With geometry, support and material fixed, displacement is linear in force. A prepared map reuses this response across contacts. A load-conditioned network recomputes the response for each new contact. We fitted 32 residual modes in axial, circumferential and radial coordinates. For each material, a regression conditioned on slenderness and relative thickness predicts mode coefficients from the coarse perturbation, fine healthy field and fine-minus-coarse healthy field. A graph on the $9\times7$ hidden markers adjusts the modes using rest geometry, inclusion site, depth and contrast. Composing modes, regression and rotations per tube leaves two matrix products per material and load. Omitting the graph gives the reduced-map control.

Training used 90 tubes, including 66 balanced across three families and eleven measured dimension configurations. Validation selected graph and load-conditioned Transolver variants, including patch deformation and a learned prior. Controls were a Transolver preparing the same modes and a geometry-conditioned reduced map. Frozen predictors faced four independently sampled contacts on each of 48 new tubes (16 per family). A separate 24 tubes calibrated simultaneous bands. Scores weight tubes equally. Known-material scores average 32 materials and four contacts. Blind forecasts use two visible readings per specimen.

\begin{table}[htbp]
\centering\small
\caption{Force queries on 48 independent simulated tubes. Perturbation error covers 6,144 known-material fields and blind error 384 future fields with inclusions. GPU times start from available FEM displacements and include host transfers, input processing and return of 128 world-coordinate perturbations. Six prespecified tubes contributed 192 timed blocks per implementation.}\label{tab:prepared}
\begin{tabular}{lrrr}\toprule
Operator & Perturbation & Blind field & Median / P95 (ms)\\\midrule
Prepared graph & 5.488\% & 3.289\% & 0.069 / 0.080\\
Load-conditioned Transolver with prior & 6.900\% & 3.292\% & 1.722 / 2.021\\
Load-conditioned Transolver without prior & 10.626\% & 3.296\% & 1.698 / 2.012\\
Prepared Transolver & 5.483\% & 3.290\% & 0.066 / 0.075\\
Reduced map alone & 5.499\% & 3.291\% & 0.067 / 0.077\\\bottomrule
\end{tabular}
\end{table}

Against the strongest load-conditioned Transolver, the prepared graph reduced perturbation error by 20.47\% (family-stratified paired 95\% interval 18.43--22.61\%) and correction-stage median/P95 latency by factors of 25.00/25.22 (Table~\ref{tab:prepared}). The blind difference was $-0.0031$ points ($-0.0082$ to $+0.0015$, noninferiority margin $+0.10$). The nominal 90\% band covered all twelve fields on 46/48 tubes. Prepared Transolver matched the graph, whose perturbation gain over the reduced map was just 0.21\% ($-0.13$ to $+0.54$\%). The active map used 2.95\,MiB per tube. Computing the coarse library, fine healthy response and reference endpoints took 18.50\,s median under concurrent local FEM runs.

\subsection{From a contact force to a wall displacement}\label{sec:force}
Table~\ref{tab:prepared} starts from available FEM displacements. Here a contact descriptor $c$ supplies the force $F(c)$ to a prepared material map $P_m$. The map combines fine homogeneous and coarse heterogeneous responses with a spatial correction and returns healthy and heterogeneous displacements at 63 hidden markers. Changing support requires a new map.

\begin{figure}[tbp]\centering\fig{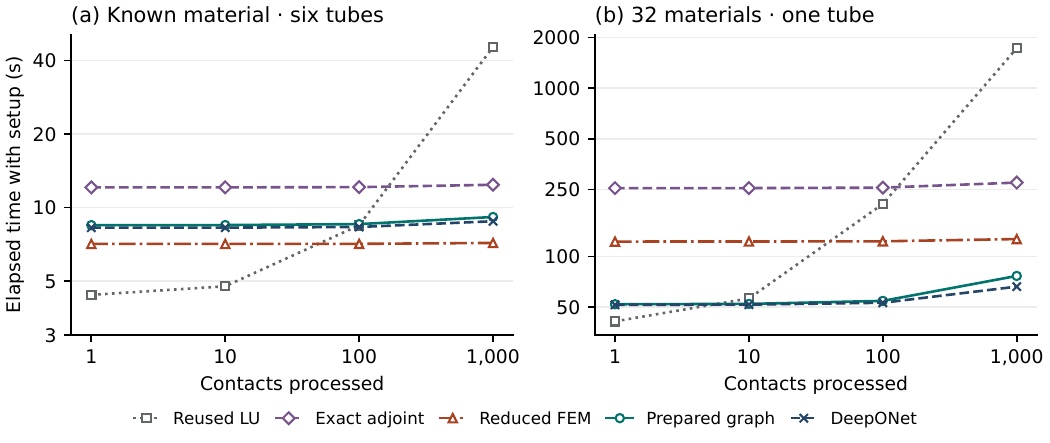}
\caption{Elapsed time on one CPU thread, including setup and the first 1, 10, 100 or 1,000 contacts. (a) Median of six tubes with one known material and original fixation. (b) All 32 materials per contact on one tube. Both axes are logarithmic. Time ranges differ between panels. See Table~\ref{tab:contacts} for accuracy.}\label{fig:contactcost}\end{figure}

Six new tubes received 1,000 independent 1\,mN contacts each under two successive fixations. One also supplied 32 materials per contact. Timings for all seven implementations include force assembly and output construction. Classical controls reused factorizations, prepared an exact output adjoint or fitted 32 modes from 48 forces. DeepONet separates input functions from evaluation coordinates \citep{deeponet}. It used 90 training tubes and a validation-selected nonlinear 64-mode branch. Table~\ref{tab:contacts} and Fig.~\ref{fig:contactcost} summarize five operators from 91,000 per-contact timings.

\begin{table}[htbp]\centering\small
\caption{Single-threaded force-to-wall computation for one known material on six tubes at their original fixation (left) and 32 materials per contact on one tube (right). Error is relative to the material-induced displacement. Total time includes preparation and 1,000 contacts.}\label{tab:contacts}
\begin{tabular}{lrr|rrr}\toprule
 & \multicolumn{2}{c}{Known material} & \multicolumn{3}{c}{32 materials}\\
Operator & Error (\%) & Hot (ms) & Error (\%) & Hot (ms) & Total (s)\\\midrule
Reused LU & 0 & 41.218 & 0 & 1686.45 & 1725.41\\
Exact output adjoint & 0 & 0.286 & 0 & 20.05 & 274.69\\
Reduced FEM, 32 modes & 3.03 & 0.065 & 2.22 & 4.77 & 126.91\\
Prepared graph & 5.04 & 0.656 & 5.66 & 23.98 & 76.64\\
DeepONet & 9.16 & 0.493 & 12.20 & 15.28 & 66.22\\\bottomrule
\end{tabular}\end{table}

With known material, the prepared graph was 62.8 times faster than reused LU per query and 4.98 times faster including setup and 1,000 contacts. Reduced FEM was faster and more accurate. For 32 materials, setup took 52.02\,s for the graph and 122.64\,s for reduced FEM, giving the graph the shorter total time despite slower queries. That ordering held on a second computer (19.57 versus 43.37\,s). Graph perturbation error ranged from 3.83 to 7.39\% across materials. DeepONet was faster for the bank but had 12.20\% perturbation error versus 5.66\% for the graph. Its total-field error was 0.27\% versus 0.52\% for reduced FEM.

\subsection{External transfer to structural brackets}\label{sec:brackets}
SimJEB supplies FEM displacements for 381 bracket designs under vertical, horizontal, diagonal and torsional loads \citep{simjeb,simjebdata}. We train new weights for this unrelated topology and score three-component relative error at 2,048 uniformly sampled vertices, weighting loads equally within design and designs equally overall. MEEC-Net used prescribed displacements at 4,096 points under vertical loading, with up to 50 training and 20 test designs \citep{shaffer}. Here the input is force.

On the original 266/57/58 design split, geometry-only errors were 45.44\% for the graph, 44.37\% for DeepONet and 96.96\% for a spatial cubic. The neural difference was inconclusive. We added elasticity on 6 and 4\,mm Cartesian grids, scaling each fully integrated trilinear voxel stiffness by the volume fraction of tetrahedra with centroids in that voxel. Interpolation constrains bolt patches and distributes pin forces and moments. The residual graph, DeepONet, pointwise model and two-coefficient linear mixture per load use both fields. Neural models share budgets and validation rules across two seeds. One empty training attachment receives zero mechanical input. All test meshes are available.

Outcomes were already exposed on the original split and three published test partitions. A further split held designers disjoint across training, validation and testing (252/52/77 designs). On every split we fitted preprocessing and models on training designs, selected checkpoints on validation designs and sealed test predictions before scoring. High-resolution test displacements were evaluator-only.

\begin{table}[htbp]\centering\small
\caption{Mean relative displacement error (\%) on the external brackets. The unit is a design, with four load cases equally weighted. The last column holds out entire designers. The graph and DeepONet rows use two seeds each.}\label{tab:brackets}
\begin{tabular}{lrr}\toprule
Predictor input and method & Design split (58) & Designer split (77)\\\midrule
Geometry only, graph & 44.59 & 47.38 \\
Geometry only, DeepONet & 45.42 & 48.56 \\
Elasticity on the 4 mm grid & 24.96 & 22.67 \\
Two grids, linear mixture & 10.83 & 9.65 \\
Two grids, pointwise residual & 9.71 & 9.28 \\
Two grids, DeepONet residual & 10.15 & 9.36 \\
Two grids, graph residual & 9.67 & 9.26 \\
\bottomrule\end{tabular}\end{table}

\begin{figure}[!htb]\centering\fig{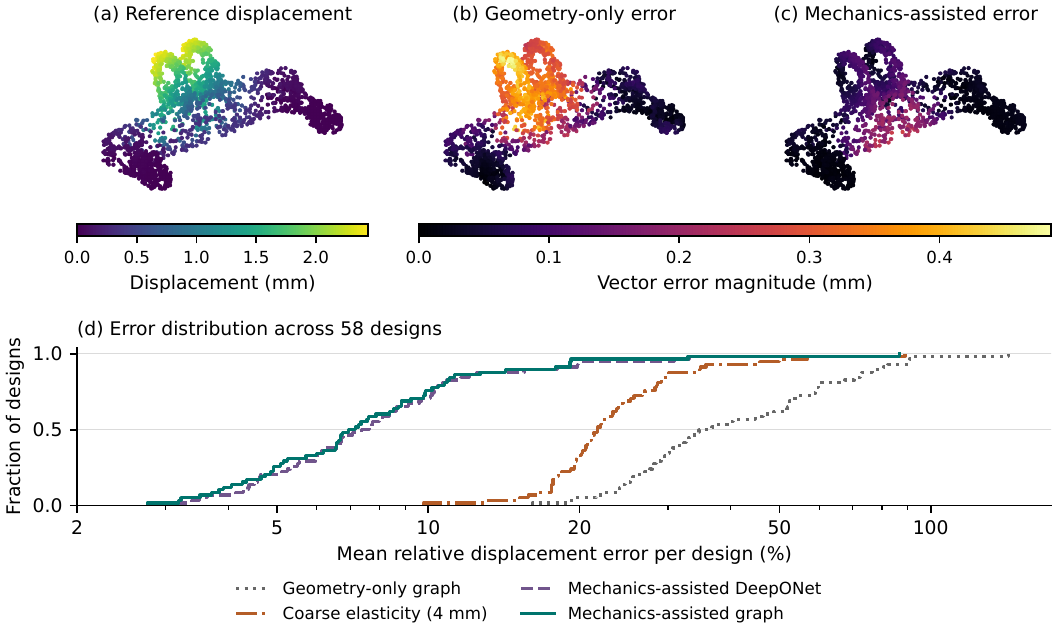}
\caption{Mechanical preparation on SimJEB. (a) Vertical-load FEM displacement magnitude at 2,048 vertices of design 551, selected by median test mesh size. (b,c) Vector errors against (a) at those vertices on a shared scale. (d) Fraction of 58 test designs below or at each threshold, using mean relative error over four loads per design. DeepONet and the graph share the coarse fields.}\label{fig:brackets}\end{figure}

The validation-selected graph gained 34.70 points over the original geometry-only DeepONet (paired 95\% interval 30.51--39.09). With equal inputs it gained 0.48 points over DeepONet (0.08--0.99), 1.16 over the linear mixture (0.03--2.44) and 0.04 over pointwise ($-0.31$ to $+0.36$, Table~\ref{tab:brackets}). Median graph error was 7.04\% and the worst decile averaged 32.50\% (Fig.~\ref{fig:brackets}). The three official partitions gave 8.85\%, 6.90\% and 8.76\%. On unseen designers the gain over equal-input DeepONet was 0.10 points (designer-clustered interval $-0.14$ to $+0.31$).

On three designs selected by mesh size, native linear tetrahedral FEM reused one sparse Cholesky factorization for four loads. Mean relative discrepancy from published fields was at most $1.5\times10^{-7}$ per design. Native preparation took 3.69--6.62 times as long as surrogate preparation including parsing, coarse solves, features and CPU correction. Both solvers can cache responses. The four loads span rank three on all 380 usable decks. Projection onto that space cut storage from 98,304 to 73,776 bytes including load coordinates, and changed primary graph error from 9.67\% to 9.60\%. Paired four- and three-column queries took 11.36 and 13.05\,$\mu$s. Across 58 designs median feature and graph-encoding times were 118 and 31\,ms. A four-column mixture took 11.52\,$\mu$s on one core. Queries combine loads at existing sites.

\section{Additional operator and transfer controls}\label{app:transfer}

\paragraph{Cost of a new geometry and a new decision.}
On eight geometries, mean cold setup plus discrete-catalogue inference took 0.314\,s for the graph, 0.793\,s for fine FEM, 0.287\,s for ridge and 0.279\,s for the unlearned baseline (one CPU thread, one warm-up, three repeats). With a prepared continuous catalogue, selection, inversion and four forecasts over 256 observation pairs on eight geometries took 4.66/5.39\,ms median/P95 for the graph and 4.67/5.40\,ms for FEM on one pinned core. Neither timing includes training, sensing or reconstruction. The graph missed the twofold classical speed target.

\paragraph{Open panels.}
For the patch operators in Section~\ref{sec:materialupdate}, $v=P^\top K_0^{-1}f$ and local force $g=-aD(v+Cg)$ give displacement $K_0^{-1}(f+Pg)$. Static condensation provides an exact control \citep{ports}. Learned elements likewise couple fine interiors to coarse exteriors \citep{parish}. On eight exposed bent tubes (512 fields), a rank-24 graph gave 1.130\% mean displacement error and 0.658 maximum relative free-force imbalance. A sensitivity loss gave 1.138\% and 0.652. Exact local condensation matched the global solution within $5.19\times10^{-14}$ mean relative error.

With known material, we initialized up to eight scaled Richardson corrections from each predicted force, targeting a 0.10 relative free-force residual. Across 1,024 withheld fields on 16 cambered panels with free transverse edges, mean full-volume errors were 0.286\% from the frozen tubular graph, 0.282\% from the nodal model, 0.297\% from rank-24 proper orthogonal decomposition (POD) and 0.780\% from zero. After adaptation on exposed panels, errors were 0.228\% for the graph and 0.172\% for POD.

\paragraph{External physical experiment.}
The Cambridge tube-law experiment tracks 15 cross-section markers under external pressure (eleven configurations, three sets of unstrained dimensions) \citep{cam}. Camera coordinates matched the deposit implementation within $4.3\times10^{-13}$\,mm. Seven one-sided markers at two pressures informed forecasts for eight later hidden markers. We froze the method after seeing the measurements and ring-model scores.

We projected repeated unloaded midsection measurements into their principal plane, interpolated the periodic radius and extruded it over the mounted length. Two-layer elasticity gives one linear mode under integrated lateral pressure and zero homogeneous graph perturbation. We prescribed the modulus or fitted nonnegative compliance to two visible steps. Controls used static shape, pressure-proportional translation and ovalisation, or the last fitted oval. Errors include camera-frame motion and registration.

\begin{figure}[htbp]\centering\fig{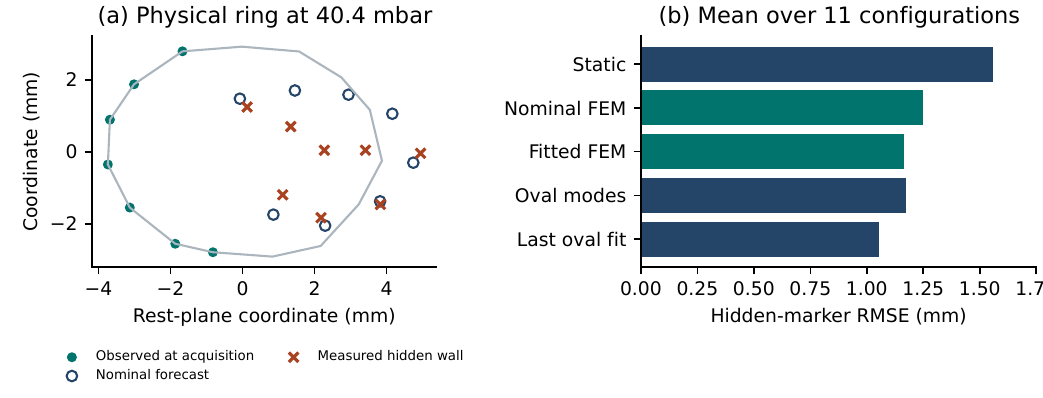}
\caption{Measured pressure-driven collapse. Left, unloaded ring and subsequent hidden-marker observations and forecasts for configuration 2. Right, one-sided vector RMSE averaged within and then across 11 configurations. All estimators use the same 368 hidden-marker requests. The nominal model uses the prescribed material. The fitted model estimates one compliance from visible data.}\label{fig:cam}\end{figure}
Across 46 later states (368/368 hidden markers), RMSE in mm was 1.250 for prescribed mechanics, 1.164 fitted, 1.558 static, 1.173 pressure-proportional ovalisation and 1.055 persistence (Fig.~\ref{fig:cam}). In the 40/46 one-sided states from ten configurations exceeding three times unloaded repeatability RMS, nominal and fitted relative errors were 79.71\% and 73.22\%. One amplitude misses the evolving oval shape. With distributed observations and their scoring mask, ovalisation gave 0.920\,mm and persistence 0.989\,mm on 362/368 requests. The deposit lacks independent contact sites and local material labels.

\paragraph{Measured completion away from tubes.}
CT tracks 117 internal markers through five indentations of one OpenCAS silicone liver phantom \citep{suwelack}. At each stage 12 visible markers informed predictions for 105 hidden markers. A graph trained on synthetic cases from the phantom's rest geometry gave 53.94\% one-sided completion error versus 54.49\% for a matched nodal network and 54.86\% for elasticity. With distributed markers, thin-plate splines gave 51.65\% and the graph 54.55\%.

\paragraph{Matched neural comparison on an additional cohort.}

A local graph and irregular-mesh Transolver \citep{transolver} shared coarse perturbations, fine healthy fields, invariant features and a four-vector output basis. Training used twelve contacts on 24 development tubes. Separate validation tubes selected checkpoints and Transolver size. Four reserved contacts on 24 prospective tubes yielded 3,072 known-material fields. Perturbation errors were 19.06\% for Transolver (133,396 parameters), 22.36\% for the graph (3,268) and 35.33\% for coarse FEM, a 14.77\% Transolver gain over the graph (13.66--15.85\%). Given mechanical inputs, GPU correction for 32 candidates and four contacts took a median 1.11\,ms for the graph and 21.56\,ms for Transolver. Both blind inverses shared two readings, likelihoods, eight homogeneous endpoints and fine-reference normalization. Total errors on 192 patched requests were 3.188\% (graph), 3.189\% (Transolver), 3.198\% (adaptive physics, $n=16$) and 5.267\% (nominal FEM, $n=32$). The graph gained 0.04\% over Transolver ($-0.69$ to $+0.72$\%).

\section*{Declaration of generative AI and AI-assisted technologies in the writing process}
The author used ChatGPT and Codex (OpenAI) for manuscript revision and assistance with analysis and plotting code. The figures were rendered from scientific data or drawn schematically as indicated in their captions. The author reviewed and edited the resulting material and takes full responsibility for the article.

\begingroup
\fontsize{9.5}{11.5}\selectfont
\setlength{\bibsep}{1pt}
\bibliographystyle{elsarticle-num}
\bibliography{references}
\endgroup
\end{document}


{\small Supplementary material for \emph{Forecasting hidden-wall deformation under material ambiguity in elastic tubes}}\par
{\Large\bfseries Supplement S1. Hidden-wall displacement and compliance under mesh refinement}\par
\'I\~nigo Cirauqui-Viloria\par

Figure S1 and Tables S1--S4 document mesh sensitivity. Figure S2 and Tables S5--S6 compare the two element orders on paired tubes. Figure S3 and Table S7 report the independent qualification. The \href{https://doi.org/10.57967/hf/11868}{reproduction release v1.2.0} includes the corresponding inputs, scores and replay code. The \href{https://huggingface.co/datasets/icirauqui/hidden-wall-deformation-reproducibility/blob/5aad8a5a1766ae21ea80faaefbe17e7f4428d6e6/README.md}{pinned release instructions} describe scoring from the supplied predictions and optional finite-element regeneration separately.

The mesh-sensitivity study (Section 5.1 of the article) uses the first replica (00) of curved family 2 in each of the four test strata. It evaluates the same 63 material markers with their original area weights. The diagnostic applies a fixed-direction Gaussian surface traction with both ends clamped. The inclusion doubles the modulus over the outer half of the specified patch.

Each row reports a mesh with n axial, 2n circumferential and n/4 radial cells. C3D4 uses linear displacements and C3D10 uses quadratic displacements on straight-sided tetrahedra. Wall values are area-weighted root-mean-square (RMS) vector magnitudes. The perturbation is the RMS of the altered field minus the healthy field, evaluated marker by marker. Compliance is the work product of the assembled nodal load and displacement (N mm). All values below are rounded for display.

\begin{center}\includegraphics[width=.94\linewidth]{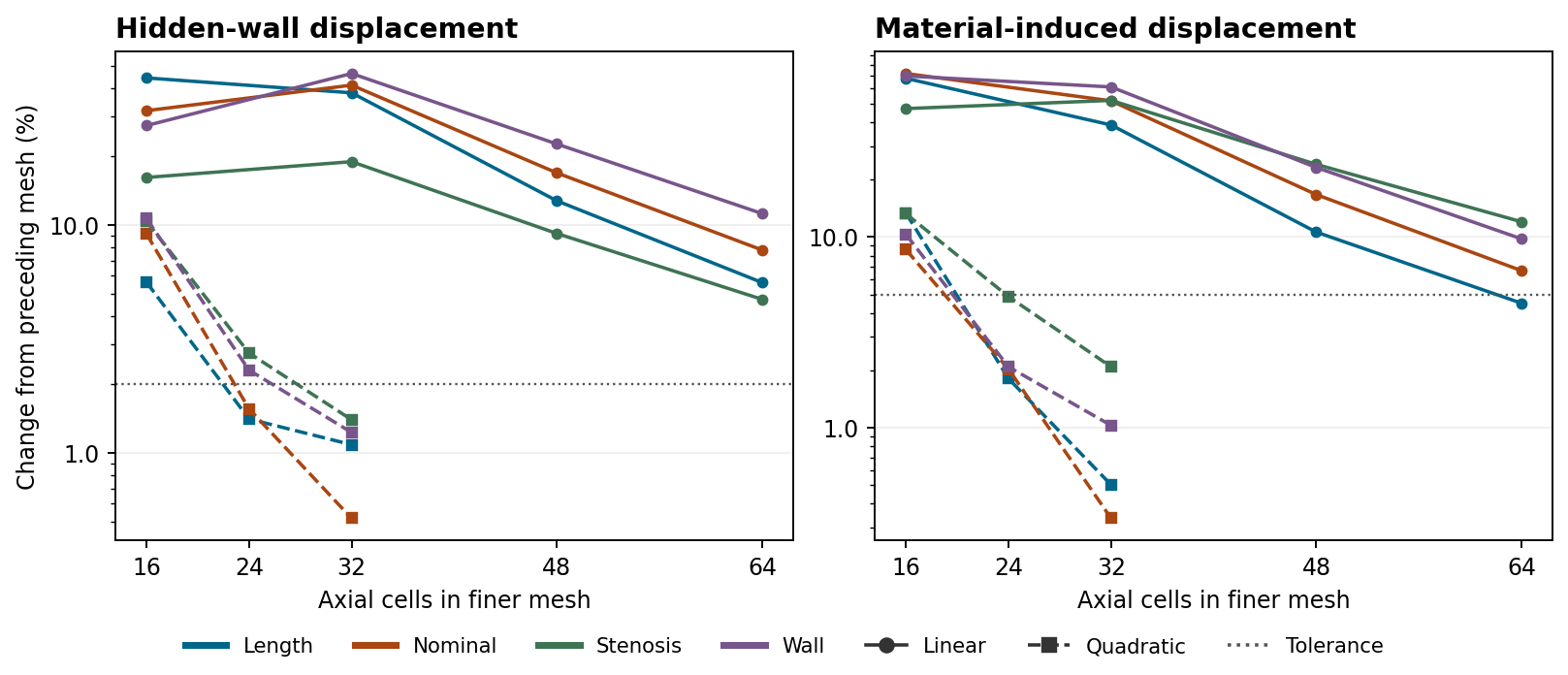}\end{center}

\textbf{Figure S1.} Each point compares successive meshes on one tube. Dotted lines mark the prespecified tolerances of 2\% for the total field and 5\% for the material-induced field. These within-order changes are one part of the refinement test. At equal displacement count (411,264 free degrees of freedom), linear and quadratic elements still differed by 7.86--17.99\% in total hidden-wall displacement across these four tubes.

\clearpage
\bigskip\textbf{Table S1.} Hidden-wall displacement and compliance for test-nominal-2-00.
\begin{center}\small\begin{tabular}{lrrrrrr}\toprule
 & & \multicolumn{3}{c}{Wall displacement (nm)} & \multicolumn{2}{c}{Compliance (N mm)}\\
\cmidrule(lr){3-5}\cmidrule(lr){6-7}
Element & $n$ & Healthy & Altered & Perturbation & Healthy & Altered\\\midrule
C3D4 & 8 & 351.632 & 346.530 & 5.569 & 2.229138e-07 & 2.162221e-07\\
C3D4 & 16 & 362.552 & 355.662 & 10.014 & 3.182719e-07 & 3.036621e-07\\
C3D4 & 32 & 389.600 & 376.468 & 17.810 & 4.017653e-07 & 3.777255e-07\\
C3D4 & 48 & 417.910 & 401.297 & 20.937 & 4.359878e-07 & 4.079895e-07\\
C3D4 & 64 & 434.601 & 416.379 & 22.260 & 4.523699e-07 & 4.225462e-07\\
C3D10 & 8 & 435.403 & 414.230 & 25.313 & 4.418878e-07 & 4.110859e-07\\
C3D10 & 16 & 458.348 & 437.532 & 24.440 & 4.725197e-07 & 4.395515e-07\\
C3D10 & 24 & 462.701 & 442.110 & 24.076 & 4.750244e-07 & 4.423962e-07\\
C3D10 & 32 & 464.580 & 443.967 & 24.074 & 4.776119e-07 & 4.449377e-07\\
\bottomrule\end{tabular}\end{center}
\bigskip\textbf{Table S2.} Hidden-wall displacement and compliance for test-wall-2-00.
\begin{center}\small\begin{tabular}{lrrrrrr}\toprule
 & & \multicolumn{3}{c}{Wall displacement (nm)} & \multicolumn{2}{c}{Compliance (N mm)}\\
\cmidrule(lr){3-5}\cmidrule(lr){6-7}
Element & $n$ & Healthy & Altered & Perturbation & Healthy & Altered\\\midrule
C3D4 & 8 & 795.709 & 783.001 & 13.290 & 5.63519e-07 & 5.478072e-07\\
C3D4 & 16 & 823.183 & 806.848 & 21.356 & 8.012772e-07 & 7.654033e-07\\
C3D4 & 32 & 869.540 & 837.383 & 42.372 & 1.046143e-06 & 9.811077e-07\\
C3D4 & 48 & 953.972 & 909.896 & 53.365 & 1.160443e-06 & 1.080841e-06\\
C3D4 & 64 & 1013.015 & 962.891 & 58.467 & 1.219218e-06 & 1.132303e-06\\
C3D10 & 8 & 1036.733 & 980.794 & 63.525 & 1.230175e-06 & 1.139912e-06\\
C3D10 & 16 & 1103.523 & 1045.179 & 65.398 & 1.304596e-06 & 1.204337e-06\\
C3D10 & 24 & 1121.411 & 1062.999 & 65.184 & 1.314582e-06 & 1.214774e-06\\
C3D10 & 32 & 1132.496 & 1073.765 & 65.359 & 1.320957e-06 & 1.221297e-06\\
\bottomrule\end{tabular}\end{center}
\clearpage
\bigskip\textbf{Table S3.} Hidden-wall displacement and compliance for test-length-2-00.
\begin{center}\small\begin{tabular}{lrrrrrr}\toprule
 & & \multicolumn{3}{c}{Wall displacement (nm)} & \multicolumn{2}{c}{Compliance (N mm)}\\
\cmidrule(lr){3-5}\cmidrule(lr){6-7}
Element & $n$ & Healthy & Altered & Perturbation & Healthy & Altered\\\midrule
C3D4 & 8 & 151.589 & 149.270 & 2.809 & 9.585198e-08 & 9.238016e-08\\
C3D4 & 16 & 163.560 & 159.371 & 5.811 & 1.41863e-07 & 1.343921e-07\\
C3D4 & 32 & 189.571 & 182.636 & 8.556 & 1.776973e-07 & 1.660816e-07\\
C3D4 & 48 & 202.962 & 195.250 & 9.208 & 1.914753e-07 & 1.782218e-07\\
C3D4 & 64 & 209.576 & 201.595 & 9.420 & 1.980859e-07 & 1.840725e-07\\
C3D10 & 8 & 211.054 & 201.959 & 10.560 & 1.910649e-07 & 1.771777e-07\\
C3D10 & 16 & 215.054 & 206.815 & 9.643 & 2.035969e-07 & 1.886306e-07\\
C3D10 & 24 & 216.503 & 208.365 & 9.521 & 2.056157e-07 & 1.905371e-07\\
C3D10 & 32 & 218.502 & 210.315 & 9.554 & 2.072506e-07 & 1.921066e-07\\
\bottomrule\end{tabular}\end{center}
\bigskip\textbf{Table S4.} Hidden-wall displacement and compliance for test-stenosis-2-00.
\begin{center}\small\begin{tabular}{lrrrrrr}\toprule
 & & \multicolumn{3}{c}{Wall displacement (nm)} & \multicolumn{2}{c}{Compliance (N mm)}\\
\cmidrule(lr){3-5}\cmidrule(lr){6-7}
Element & $n$ & Healthy & Altered & Perturbation & Healthy & Altered\\\midrule
C3D4 & 8 & 259.525 & 255.937 & 3.716 & 1.174148e-07 & 1.146006e-07\\
C3D4 & 16 & 280.167 & 276.217 & 4.519 & 1.578514e-07 & 1.528206e-07\\
C3D4 & 32 & 281.946 & 277.907 & 5.936 & 1.928989e-07 & 1.851776e-07\\
C3D4 & 48 & 284.188 & 279.852 & 7.102 & 2.080356e-07 & 1.989669e-07\\
C3D4 & 64 & 286.467 & 281.869 & 7.831 & 2.156931e-07 & 2.05927e-07\\
C3D10 & 8 & 279.940 & 272.477 & 11.376 & 2.258955e-07 & 2.143926e-07\\
C3D10 & 16 & 287.555 & 281.463 & 10.309 & 2.296095e-07 & 2.180404e-07\\
C3D10 & 24 & 289.579 & 283.855 & 9.860 & 2.295636e-07 & 2.182424e-07\\
C3D10 & 32 & 291.229 & 285.655 & 9.677 & 2.299478e-07 & 2.187405e-07\\
\bottomrule\end{tabular}\end{center}\clearpage

\textbf{Two readings under a changed mechanical reference.} One replica per geometry stratum and curved family gave 12 tubes. Each retained its original three material states, two optical acquisition loads, four future loads, common camera and pixel-noise draw. A C3D10 mesh with 32 or 40 axial cells, twice as many angular cells and four radial cells supplied both readings and targets. The same lower-order inference libraries were used on both references.

The P1 nodal forces $f$ define a piecewise linear traction through $Mt=f$, with the consistent mass matrix $M$ on the original exterior triangles. Integrating $t$ against the quadratic trace preserves its work on every original linear trace. The cells inherited the original material labels. Distal releases extend over the positive support of the original released nodal basis functions. These choices specify how the original discrete loading and support continue onto the richer displacement space.

\textbf{Table S5.} Mean area-weighted vector RMS errors on 144 future fields. Material-relative error uses the 96 altered fields. Coverage requires all twelve fields on one tube to lie within their radii. The mean diameter is twice the mean calibrated RMS radius. The two-reading multiplier was also applied at lower budgets without recalibration.

\begin{center}\small\begin{tabular}{llrrrrr}\toprule

Reference & Readings & Total (\%) & RMS ($\mu$m) & Material (\%) & Covered & Diam. ($\mu$m) \\

\midrule

C3D4 & 0 & 4.96 & 0.095 & 100.1 & 12/12 & 0.708 \\

C3D4 & 1 & 4.96 & 0.094 & 105.5 & 9/12 & 0.458 \\

C3D4 & 2 & 3.10 & 0.053 & 104.0 & 11/12 & 0.321 \\

C3D10 & 0 & 49.58 & 1.409 & 100.0 & 0/12 & 0.708 \\

C3D10 & 1 & 41.98 & 1.192 & 236.0 & 0/12 & 0.235 \\

C3D10 & 2 & 41.29 & 1.180 & 281.8 & 0/12 & 0.132 \\

\bottomrule\end{tabular}\end{center}

\begin{center}\includegraphics[width=.90\linewidth]{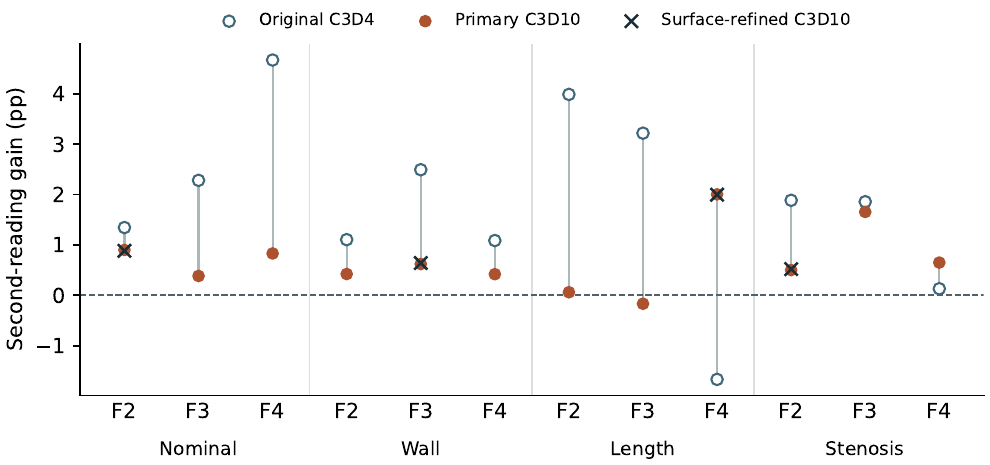}\end{center}

\textbf{Figure S2.} Gain from the second reading on the same tubes and loads. Positive values indicate smaller relative total-field error. Each tube contributes twelve forecasts. Crosses show the additional surface refinement on the four prespecified tubes. F2--F4 denote curved geometry families.

\clearpage\textbf{Refinement on the acquisition and forecast loads.} On one tube per stratum we doubled radial subdivisions from two to four, then doubled both surface subdivisions at four radial subdivisions. Table S6 gives maxima across the six loads and relevant material states. The optical discrepancy uses the RMS pixel difference divided by the original likelihood standard deviation. Each wall comparison uses the same 63 material markers and the finer field as denominator.

\textbf{Table S6.} Successive C3D10 changes. The limits were 2\% for total field and compliance, 5\% for the material effect and one noise standard deviation for optical readings.

\begin{center}\small\begin{tabular}{llrrrr}\toprule

Tube & Step & Total (\%) & Material (\%) & Work (\%) & Optical (SD) \\

\midrule

Nominal & Radial & 0.18 & 0.28 & 0.20 & 0.01 \\

Nominal & Surface & 1.55 & 2.45 & 1.01 & 0.07 \\

Wall & Radial & 0.10 & 0.27 & 0.18 & 0.01 \\

Wall & Surface & 1.28 & 1.52 & 0.81 & 0.15 \\

Length & Radial & 0.32 & 0.58 & 0.32 & 0.02 \\

Length & Surface & 1.42 & 2.12 & 0.89 & 0.05 \\

Stenosis & Radial & 0.24 & 0.27 & 0.25 & 0.02 \\

Stenosis & Surface & 1.31 & 1.95 & 1.36 & 0.19 \\

\bottomrule\end{tabular}\end{center}

The first reading gained 7.60 [6.92, 8.23] percentage points on the primary quadratic reference and the two readings together gained 8.29 [7.51, 8.98] (paired 95\% intervals). The mean total errors with two readings were 41.5099\% for the fitted filter and 41.5116\% for its degree-34 control under the shared posterior. The nominal homogeneous field gave 51.79\% error. On the four sentinels every refinement threshold was met. The largest surface-refinement change in the second-reading gain was 0.021 percentage points. One of the 12 primary tubes (length, F3) worsened by 0.17 points. The reference comparison is conditional on the stated continuation of the discrete loads and support.

\clearpage
\textbf{High-fidelity qualification.} We drew 36 new geometries across the same three families and four strata. Twelve calibration and 24 confirmation tubes supplied one homogeneous and two heterogeneous specimens each. Nominal inference used 24-ring C3D10 responses on four radial subdivisions. On each tube the independent evaluator used 40 or 48 rings, the same four radial layers, patch centres and footprints outside the catalogue, unknown support and thickness, nominal modulus shifted by up to 2\%, and smooth radius variations up to 0.3\%. Six triangle quadrature points integrate the prescribed vector traction at acquisition and forecast contact loads. Resultants, moments and reactions were checked on the retained four paired diagnostics. The P1-equivalent and continuous traction choices were assessed separately. On two new calibration tubes, refinement changed total displacement by less than 0.90\%. The material-only changes of 9.22\% and 16.12\% fell to 1.64\% and 3.15\% when the prior patch footprint was held fixed by parametric cell averaging.

For contrast $r$ set $a=(r-1)/(r+1)$ and $X=2K_0^{-1}K_p-I$. With $q=K_p u_0$, propagate $v_k=T_k(X)K_0^{-1}q$ using a single reference factorization. The updated field is $u_0-2a(I+aX)^{-1}v_0$. Expanding in the Chebyshev basis gives coefficient $-2a/\sqrt{1-a^2}$ at degree zero, and $-4ac^k/\sqrt{1-a^2}$ at degree $k>0$, where $c=-a/(1+\sqrt{1-a^2})$. The absolute scalar tail is bounded by $2.1\times10^{-12}$ at degree 16 over $r\in[0.45,2.15]$ in the homogeneous energy norm per unit seed norm. Direct factorization verifies the first stiff patch for every prepared geometry. The Galerkin control solves a three-dimensional system per contact using the same first three basis vectors. The fitted and fixed filters use those vectors with polynomial weights.

\begin{center}\includegraphics[width=\linewidth]{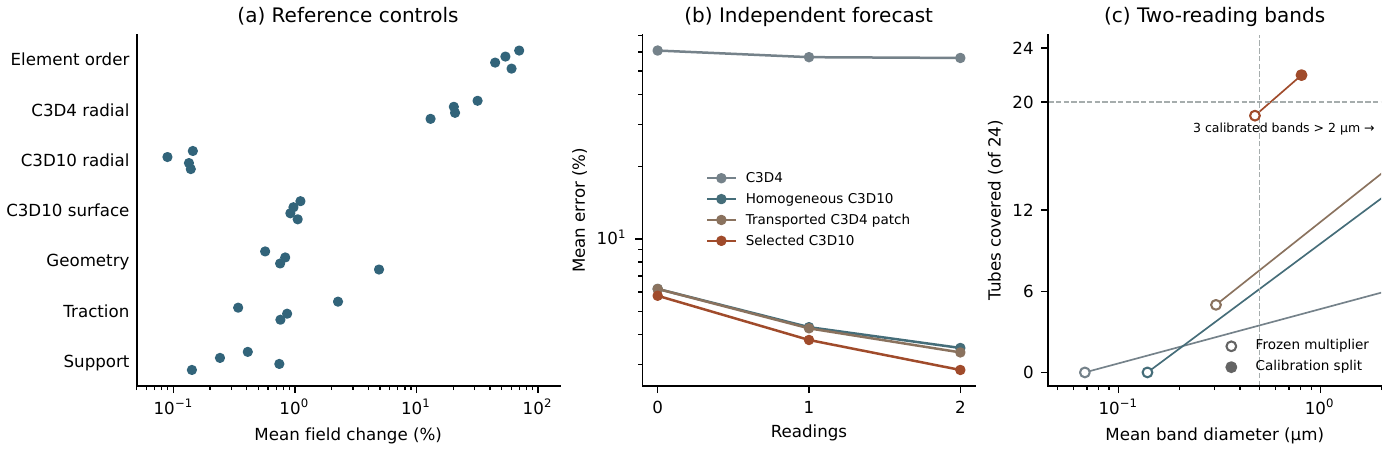}\end{center}
\textbf{Figure S3.} Qualification of the shared reference. Panel (a) changes a single mechanical component on four diagnostic tubes. Panel (b) compares the linear inference baseline with quadratic alternatives on the 24 independent confirmation tubes. In panel (c), open markers apply the historical multiplier and filled markers apply the independent calibration. The dashed lines mark the registered coverage and width criteria.
\clearpage
\textbf{Table S7.} Two-reading qualification and cost controls. Material error is relative to material-induced displacement. The known-material column compares nominal catalogue endpoints to converged C3D10 reanalysis. The energy score is the mean joint four-load energy score. Queries evaluate all four wall fields for a new contrast at an existing site. Preparation excludes homogeneous anchor construction, which is common to these patch-update controls.
\begin{center}\small\begin{tabular}{lrrrrr}\toprule
Method & Material (\%) & Known (\%) & Energy ($\mu$m) & Prep. (s) & Query ($\mu$s) \\ \midrule
Fitted & 103.4 & 0.487 & 0.0760 & 30.03 & 2.01 \\
Fixed polynomial & 103.3 & 0.647 & 0.0760 & 30.03 & 2.23 \\
Chebyshev degree 2 & 103.4 & 0.564 & 0.0761 & 30.03 & 2.66 \\
Reduced Galerkin & 103.3 & 0.329 & 0.0762 & 30.03 & 5.80 \\
Converged & 103.4 & 0.000 & 0.0762 & 119.19 & 4.50 \\
\bottomrule\end{tabular}\end{center}

The selected band multiplier was the largest tube nonconformity among 12 calibration tubes. Two-reading mean energy score was 0.0760\,$\mu$m. The original multiplier gave mean diameter 0.474\,$\mu$m and covered 19/24 tubes. A gate set to the calibration upper quartile of normalized optical incompatibility retained 13/24 confirmation tubes. This is an optical model check under the assumed sensor noise. Retained and rejected tubes had 2.85\% and 2.84\% total error. This gate did not improve selective prediction. Descriptively, using one common multiplier would require at least 0.532\,$\mu$m mean diameter to cover 20/24 test tubes. A 0.50\,$\mu$m mean diameter covers 19/24. The test data set no operating threshold.